\documentclass[letterpaper,twocolumn,10pt]{article}
\usepackage{usenix}

\usepackage{tikz}
\usepackage{amsmath}

\usepackage{tikz}
\usepackage{amsmath}
\usepackage{amsfonts}

\usepackage[english]{babel}
\usepackage{blindtext}
\usepackage{lipsum}
\usepackage{amsthm}
\usepackage[ruled,vlined,linesnumbered]{algorithm2e}

\usepackage{filecontents}
\usepackage{xspace}
\usepackage{xcolor} 
\usepackage{hhline}
\usepackage{multirow}
\usepackage{soul}
\usepackage{subcaption}

\usepackage{array}
\usepackage{mathtools}
\usepackage{comment}

\usepackage{pifont}
\usepackage{marvosym}

\usepackage{csquotes}
\usepackage[font=itshape]{quoting}
\usepackage{filecontents}

\usepackage{tikz}
\newcommand{\highlight}[2]{\colorbox{#1!17}{#2}}

\usepackage{todonotes} % for comments

\newcommand{\cameraold}[1]{{\color{red}{}}}

\newcommand{\jc}[1]{{\color{red}{\footnotesize [JC: #1]\xspace}}}

\newcommand{\ignore}[1]{{\xspace}}
\newcommand{\REMOVE}[1]{}

\newcommand{\fillme}{{\bf XXX}\xspace}

\newcounter{packednmbr}
\newenvironment{packedenumerate}{\begin{list}{\thepackednmbr.}{\usecounter{packednmbr}
\setlength{\itemsep}{0.5pt}\addtolength{\labelwidth}{-4pt}\setlength{\leftmargin}{2.5ex}\setlength{\listparindent}{\parindent}\setlength{\parsep}{1pt}\setlength{\topsep}{2pt}}}{\end{list}}
\newenvironment{packeditemize}{\begin{list}{$\bullet$}{
\setlength{\itemsep}{0.5pt}\addtolength{\labelwidth}{-4pt}\setlength{\leftmargin}{2.5ex}\setlength{\listparindent}{\parindent}\setlength{\parsep}{1pt}\setlength{\topsep}{2pt}}}{\end{list}}

\newcommand{\tightcaption}[1]{\vspace{-0.4cm}\caption{{\normalfont{\textit{{#1}}}}}\vspace{-0.55cm}}

\newcommand{\tightsection}[1]{\vspace{-0.35cm}\section{#1}\vspace{-0.1cm}}

\newcommand{\tightsubsection}[1]{\vspace{-0.3cm}\subsection{#1}}\vspace{-0.1cm}

\newcommand{\eg}{{\it e.g.,}\xspace}
\newcommand{\ie}{{\it i.e.,}\xspace}

\newcommand{\myparashort}[1]{\smallskip\noindent{\bf {#1}}~}
\newcommand{\mypara}[1]{\vspace{0.1cm}\noindent{\bf {#1}:}~}

\newcommand{\myparaq}[1]{\vspace{0.1cm}\noindent{\bf {#1}?}~}

\newcommand{\rate}{{DMR}\xspace}

\newcommand{\name}{\textsf{Grace}\xspace}
\newcommand{\newae}{\textsf{Grace-AE}\xspace}

\begin{document}
%-------------------------------------------------------------------------------

%don't want date printed
\date{}

%\title{\scalebox{0.9}{\Large \name: Loss-Resilient Real-Time Video Communication Using Data-Scalable Autoencoder}}
%\title{\scalebox{0.97}{\Large \name: Real-Time Video Communication Using Data-Scalable Autoencoders}}
\title{\scalebox{0.89}{\Large \name: Balancing Quality and Tail Delay in Real-Time Video via Data-Scalable Autoencoders}}

%for single author (just remove % characters)
% \author{
% $\text{Yihua Cheng}^{1}$, $\text{Anton Arapin}^{2}$, $\text{Ziyi Zhang}^{1}$, $\text{Qizheng Zhang}^{1,3}$, $\text{Hanchen Li}^{1}$, $\text{Nick Feamster}^{1}$, $\text{Junchen Jiang}^{1}$ \\
% $^{1}$\{yihua98,ziyizhang,qizhengz,lihanc,feamster,junchenj\}@uchicago.edu \xspace \xspace
% $^{2}$anton.arapin@gmail.com \\
% $^{1,2}$The University of Chicago, $^{3}$Stanford University
% }

\author{
$\text{Yihua Cheng}^{1}$, $\text{Anton Arapin}^{1}$, $\text{Ziyi Zhang}^{1}$, $\text{Qizheng Zhang}^{2}$, $\text{Hanchen Li}^{1}$, $\text{Nick Feamster}^{1}$, $\text{Junchen Jiang}^{1}$ \\
$^{1}$The University of Chicago, $^{2}$Stanford University
}

\maketitle

%!TEX root = ../main.tex
%!TEX spellcheck = en_US

\begin{abstract}
%Your abstract text goes here. Just a few facts. Whet our appetites.
%Not more than 200 words, if possible, and preferably closer to 150.

Across many real-time video applications, we see a growing need (especially in long delays and dynamic bandwidth) to allow clients to decode each frame once {\em any} (non-empty) subset of its packets is received and improve quality with each new packet.
We call it {\em data-scalable} delivery. 
%This is pressingly needed .
Unfortunately, existing techniques (\eg FEC, RS and Fountain Codes) fall short---they require either delivery of a minimum number of packets to decode frames, and/or pad video data with redundancy in anticipation of packet losses, which hurts video quality if no packets get lost. 
This work explores a new approach, inspired by recent advances of neural-network autoencoders, which make data-scalable delivery possible. 
We present \name, a concrete data-scalable real-time video system.
With the {\em same} video encoding, \name's quality is slightly lower than traditional codec without redundancy when no packet is lost, but {\em with each missed packet, its quality degrades much more gracefully} than existing solutions, allowing clients to flexibly trade between frame delay and video quality.
\name makes following contributions.
It trains new custom autoencoders to balance compression efficiency {\em and} resilience against a wide range of packet losses; and 
it uses a new transmission scheme to deliver autoencoder-coded frames as individually decodable packets.
We test \name (and traditional loss-resilient schemes and codecs) on real network traces and videos, and show that while \name's compression efficiency is slightly worse than heavily engineered video codecs, it significantly reduces tail video frame delay (by 2$\times$ at the 95th percentile) with the marginally lowered video quality.

\end{abstract}

%\input{sections/abstract_v3}
%\input{sections/abstract_v2}

%!TEX root = ../main.tex
%!TEX spellcheck = en_US

\tightsection{Introduction}

%Real-time videos are as important and popular as traditional streaming videos. 
With applications ranging from video conferencing to emerging IoT services and cloud-based gaming~\cite{webrtc-gaming-1,webrtc-gaming-2,webrtc-vr-1,webrtc-vr-2,webrtc-iot-1,webrtc-iot-2}, real-time videos in Chrome over WebRTC (one of the most popular real-time video frameworks) grew 100x during 2020~\cite{blum2021webrtc}, and one global industry survey found that services based on real-time videos grew from less than 5\% to over 28\% between 2020 and 2021~\cite{video-report-2020,video-report-2021}.
% ~\cite{\url{https://www.wowza.com/wp-content/uploads/Streaming-Video-Latency-Report-Interactive-2020.pdf},??}.
% https://f.hubspotusercontent20.net/hubfs/229276/2021%20Low%20Latency%20Report/streaming-video-latency-report-interactive-2021.pdf

With this increasing demand for real-time videos, it is crucial to deliver the video frames smoothly and with decent quality under a wide range of network conditions\footnote{This problem is so profound and urgent that serious efforts are finally being made by ISPs to offer active queue management for real-time traffic~\cite{l4s}, marking  a shift in focus from bandwidth to tail delay~\cite{latency-report-apnic}, but wide deployment is still a long way off. Our proposed solution is complementary to this trend and can benefit from it once deployed (see \S\ref{sec:eval}).}, including
long latency and highly dynamic network jitters, causing intermittent bursts of {\em packet losses}. 
Importantly, in this work, we use packet losses to refer to both randomly dropped packets {\em and} those delayed (due to congestion, path fluctuation, or retransmission) and thus arriving after when the frames they belong to are supposed to be decoded (\eg 40ms after decoding the last frame if the video is 25fps).
%{\em long latency}, {\em high packet jitters} and {\em intermittent bursts of packet losses}.
%Importantly, in real-time video, packet losses include {\bf not only} randomly dropped packets, but also those delayed by congestion, path fluctuation, or retransmission and arriving after when the frames they belong to are supposed to be decoded (\eg 40ms after decoding the last frame if the video is 25fps)\footnote{In lack of a better term, we call them {\em packet losses}.}.
Recent measurements show that high packet jitters and losses exist widely in real-time video applications~\cite{jansen2018performance,concerto,latency-report-apnic,latency-report-iab}; \eg a drop of bandwidth from 3Mbps to 500Kbps can cause 75\% of packets missing their playback time and raise the RTT by 6$\times$.
%one study found that in \fillme sessions, packet loss rates can be up-to \fillme\% for \fillme\% frames.\jc{YIHUA, please fill the numbers}
With long RTT, retransmission is often too late since the jitter buffer is only tens of milliseconds; \eg resending packets on a path with 100ms RTT will delay or skip at least three frames in a 25fps video stream (\ie a new frame every 40ms).

\begin{comment}
Since real-time video clients have a very shallow jitter buffer (\fillme), a packet is effectively lost as long as it is delayed long enough (due to retransmission, congestion, path fluctuation, etc) to affect the decoding of a frame (and subsequent frames).\footnote{It is important that in real-time video, packet losses {\bf not only} include packets randomly dropped by the network, but also those that are delayed due to congestion, path fluctuation, or retransmission and arrive later than when the frame it belongs to is supposed to be decoded (\eg 40ms after the last frame was decoded if the video is 25fps). For lack of a better term, we call them packet losses, and the packet loss rates can thus easily be as high as \fillme\%, as shown in recent measurements~\cite{??,??}}
For instance, retransmitting packets on a path with 100ms RTT could delay (or skip) at least three frames in a 25fps video stream (\ie a new frame every 40ms).
Prior measurements have confirmed that such long RTT is common in video conferencing and that high packet jitters and losses are not rare, due to reasons including bursty cross-traffic, packet reordering, and variation of encoded frame sizes.
\end{comment}

%\jc{mention the need for high frame rate, i.e., can't skip frames!}

\begin{figure}[t!]
    \centering
         \centering
        \includegraphics[width=0.85\linewidth]{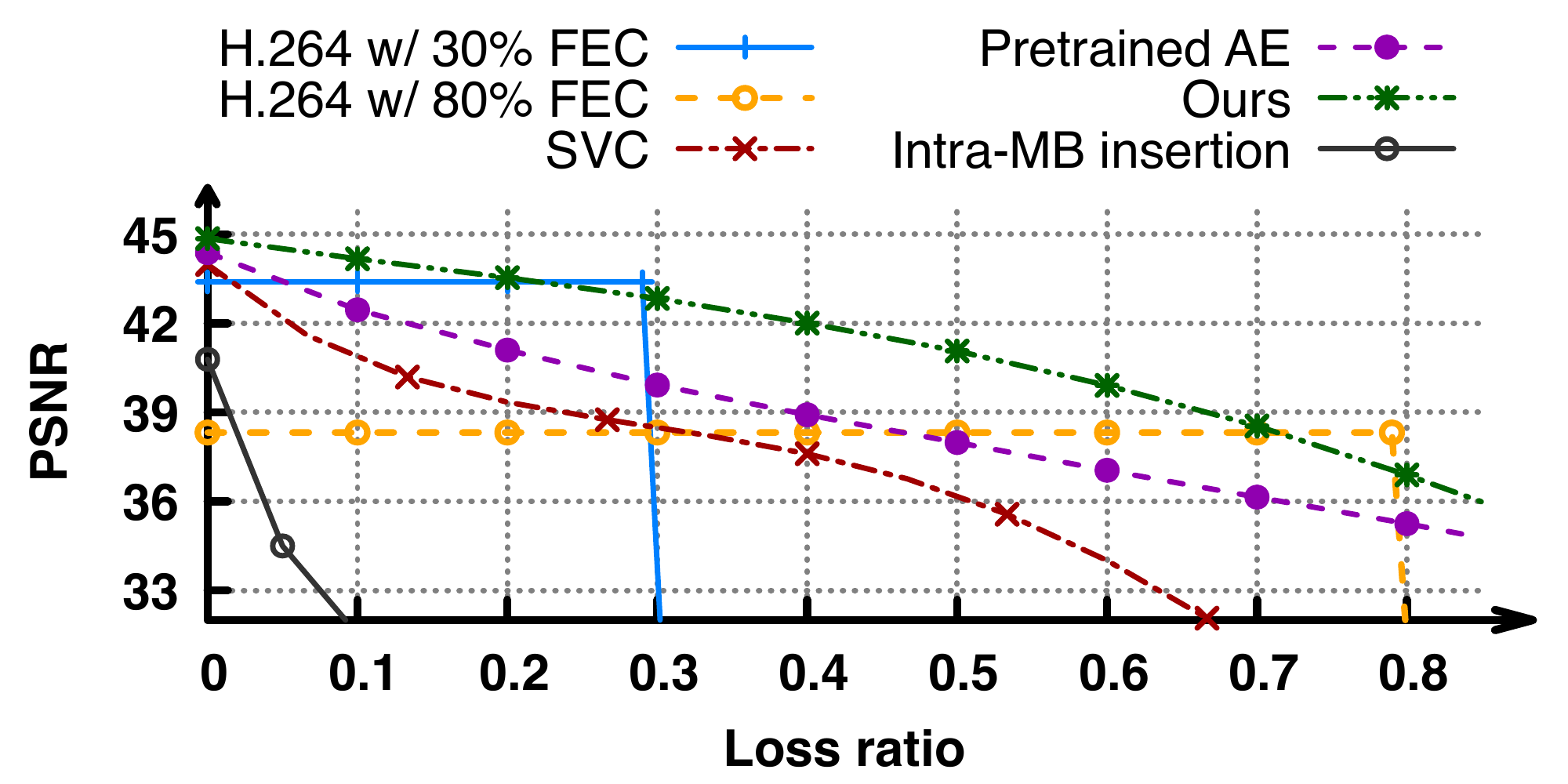}
         \label{fig:}
    \tightcaption{Comparing different loss-resilient schemes when the actual encoded bitrate (not target bitrate) is set at 3.6 Mbps. (FEC bits are included in the encoded bitrate.)
    Quality of \name's autoencoder degrades gracefully with a higher packet loss rate. 
%    achieves a smoother quality degradation as packet loss rate increases.
    More results can be found in \S\ref{sec:eval}.
    %\jc{YIHUA, remember to add a line of 264 without loss, and remove the part of h264 when it's undecodable}\jc{set R-D optimization to "intra-MB insertion"}
    }
    \label{fig:intro-example}
    % \vspace{-0.5cm}
\end{figure}

%To flexibly balance video quality and video delay, ideally a client should be able to decode the next frame when {\em any} non-empty subset of its packets is received and still obtain decent quality given the amount of data received. 
Many prior techniques have been proposed to tolerate packet losses, but they either require retransmission or trade low quality for higher loss resilience. 
%
%An ideal client should therefore tolerate packet losses be able to decode the next frame when {\em any} non-empty subset of its packets is received and still obtain decent quality given the amount of data received. 
%Unfortunately, existing loss-resilient schemes fall short of this expectation.
%, especially in links with long propagation RTTs and high packet jitters (\eg transient congestion and packet drops). 
For instance, Forward Error Correction (FEC) requires a minimum fraction of packets to be reliably received before a frame is decodable, and to lower this minimum fraction, more redundancy must be added, forcing the sender to encode a frame in lower quality before sending the packets out. 
As a result (illustrated in Figure~\ref{fig:intro-example}), the video quality will be low if packet losses are overestimated, or delay will be high if packet losses are underestimated. 
Similar tradeoffs between quality and delay (loss resilience) manifest themselves in other schemes (\eg SVC, intra-MB insertion) as well (\S\ref{subsec:abstractions}). 

%To tolerate a higher fraction of packets being missed when decoding, more redundancy must be added {\em before} packets transmission. 
%Thus, choosing an appropriate level of redundancy is known to be difficult under high packet jitters~\cite{??}.
%As illustrated in Figure~\ref{fig:intro-example}, too much redundancy would underutilize bandwidth (\ie low video quality if all packets arrive on time), while too little redundancy would drive the receiver to wait for (or resend) more packets. 
%In short, the same coding cannot achieve high quality under both no packet loss and high fraction of packet loss.
%As elaborated in \S\ref{??}, other schemes (\eg SVC, inter-MB insertion) also must deliver at least certain number of packets for a frame to be decodable, and/or pad video data with non-trivial redundancy in anticipation of certain packet loss rate, which hurts video quality if no packets turn out lost.

This paper presents the design and implementation of \name, a new real-time video system based on the emergent neural {\em autoencoders}.
%To make data-scalable autoencoders practical, we present the design and implementation of \name, an end-to-end real-time video system.
Figure~\ref{fig:intro-example} shows an example of \name's loss resilience: 
if no packet is lost, a frame can be decoded at a similar quality to classic codecs (\eg H.264), and with more packet losses, \name's quality degrades more {\em gracefully} than various baselines (explained in \S\ref{sec:eval}).

At its core, \name's new autoencoder offers a new abstraction, called {\em data-scalable delivery}:
the same encoding of a frame can be decodable once {\em any} non-empty subset of its packets are received (as opposed to a minimum set of packets required by FEC, RS or fountain codes) and achieve higher quality with each new packet received.\footnote{At first glance, this may sound like scalable video coding (SVC), whose quality increase with throughput. However, SVC is not exactly data-scalable, because it imposes a strict hierarchy among packets and a missing packet may disproportionally render multiple quality layers undecodable. In practice, SVC often employs extra loss-protection like FEC but only to protect its base layer (known as unequal error protection).}
%the decoded video quality, under no packet losses, should be similar to a classic codecs (\eg H.264, H.265) and degrade {\em gracefully} with more lost packets. 
Therefore, a sender does not need to predict packet losses; instead, the receiver can flexibly choose when to decode a frame once any packets have arrived while enjoying decent video quality.

\name customizes autoencoders because of their empirical smooth relationships between video quality and the perturbations on the coded data (\eg packet losses), which naturally arise from autoencoders' neural network structure (elaborated in \S\ref{subsec:understanding}).
However, off-the-shelf autoencoders do not yet have a graceful quality degradation under packet losses.

%of autoencoders makes them a natural fit for data-scalable delivery, thanks to the autoencoder's smooth relationship between video quality and the perturbations on the coded data (elaborated in \S\ref{sec:understanding})

\name's contribution is three-fold. 
First, we show that by simulating packet losses during autoencoder training (\S\ref{sec:training}), it is possible to train autoencoders that learn not only how to compress frames efficiently but also how to code them in ways resilient to lost packets.\footnote{\name's autoencoder is joint-source-channel coder (see \S\ref{sec:related}).}
%, in a similar spirit as~\cite{??}, which also uses autoencoders, but to code images (rather than video) to correct bit-level noises (rather packet losses).}
%We also discuss design choices regarding how packet losses should be simulated in end-to-end training and how to train autoencoders to work well on various packet loss rates. 
%, and
%what packet losses should be revealed during autoencoder training in order to balance quality at high rates and low rates of packet losses.

Second, \name presents a framework to stream autoencoder output over a {\em sequence} of frames (\S\ref{sec:delivery}). 
\name uses custom strategies to encode sequences of I-frames and P-frames and allow a receiver to determine when to decode a frame anytime after receiving the first packet, in order to balance frame delay and decoded video quality even when packet losses last for multiple frames. 
%of frame scheduling (when to decode a frame) to allow the receiver to balance frame delay and decoded video quality  are received and decoded in a way that balances frame delay and quality.
\name can work on existing congestion control algorithms, which decide how fast packets should be sent, while \name decides {\em what} packets should be sent.

Third, autoencoders are not as optimized (in speed and coding efficiency) as heavily engineered traditional codecs. 
To make it practical, \name provides optimizations to speed up the encoding/decoding, reduce compute and memory overheads, and adapt encoding bitrates under dynamic bandwidth (\S\ref{subsec:optimize}).
Importantly, \name does {\em not} improve coding efficiency (higher quality with fewer bits), which most heavily engineered video codecs strive to optimize. 
Indeed, \name's video quality is still slightly lower than H.265 at the same bitrate. 
%That said, \name enables data-scalable delivery to achieve better quality with lower delay in high jitter networks.

We are not the first to apply autoencoders to data communication. 
Autoencoders have already been used in wireless and data communication to compress images, video files, and signals~\cite{erpek2020deep}. Notably, a recent work also uses custom autoencoders to obtain highly efficient scalable video coding (SVC)~\cite{swift}. 
%Many aspects of \name are also inspired by prior work. 
That said, \name, to our best knowledge, is the first to develop custom autoencoders that realize data-scalable video delivery.
The contribution of \name is an end-to-end real-time video delivery framework, which tackles various challenges arising from training and using data-scalable autoencoder in real-time video communication.

We test \name with traditional loss-resilient schemes and codecs on real and synthetic network traces, as well as various genres of videos.
Our results show that while \name's compression efficiency is slightly worse than heavily engineered video codecs, it significantly reduces {\em tail} frame delay (by 2$\times$ at the 95th percentile when prior work has to retransmit packets or skip frame), with the marginally lowered video quality.
To put it in perspective, \name can significantly reduce the chance that frames arrive after user-perceivable delay (\eg 200ms for video conferencing) and improve playback smoothness (higher frame rate).

\textbf{This work does not raise any ethical issues.}

%\input{sections/intro_v4}
%\input{sections/intro_v3}
%\input{sections/intro_v2}

%!TEX root = ../main.tex
%!TEX spellcheck = en_US

\tightsection{Loss tolerance for real-time video}

%\begin{figure*}[t!]
%    \centering
%         \centering
%        \includegraphics[width=0.92\linewidth]{figs/primitive_table_new.pdf}
%    \caption{Abstractions of different loss-resilient schemes. The differences are: 
%    (1) unlike redundancy-based schemes, data-scalable coding does not depend on prediction of packet loss rate; and 
%    (2) unlike retransmission-based or redundancy-based schemes which must retransmit lost packets until some {\em predefined set} or {\em minimum number} of packets are received, data-scalable delivery can decode a frame with any non-empty set of packets, and the quality gradually improves with every new packet.}
%    \label{fig:primitive-table}
%\end{figure*}

% Real-time video, it is crucial that enough data is timely received for each frame, so that the high-quality frames can be rendered at a fixed interval (\eg every 50ms for 20fps). 
% However, as most packet networks have occasional congestion, delay jitter and packet losses, not all packets of a frame may arrive timely.
% Retransmitting missing packets is also inefficient if the round trip time is long.
% Therefore, loss-resilient schemes are required to ensure that a frame can be decoded at a high quality even if a fraction of its packets have not arrived when it is supposed to be decoded, \ie non-zero deadline-missing rate or \rate. 

Many recent studies (\eg~\cite{macmillan2021measuring,jansen2018performance,concerto,moulay2018experimental}) have shown that intermittent congestion, packet jitter, and packet drops widely exist and often cause long tail delay and low frame rate in real-time video applications. 
Consequently, a client may not be able to receive all packets of a frame before the frame is due for decoding (\eg 40ms after the last frame was decoded if the video is 25fps).
Thus, the ability to decode a frame, despite packet losses (including those dropped or those delayed by congestion), is pressingly needed. 

\begin{comment}
Thus, the ability to tolerate packet losses\footnote{In this work, a packet loss can occur even if the packet is not dropped by the network; a packet is lost if it is delayed (due to congestion, path fluctuation, etc) and arrives after the receiver decodes the frame it belongs to.} is pressingly needed to allow a real-time video client to decode and render the next frame when only a subset of its packets are received.
\end{comment}

\tightsubsection{Prior loss-resilience schemes}
\label{subsec:abstractions}

%This section revisits three general approaches to loss resilience taken by prior work.  

%\subsection{Prior Loss-Resilient Schemes}

\begin{figure}[t!]
    \centering
         \centering
        \includegraphics[width=0.999\linewidth]{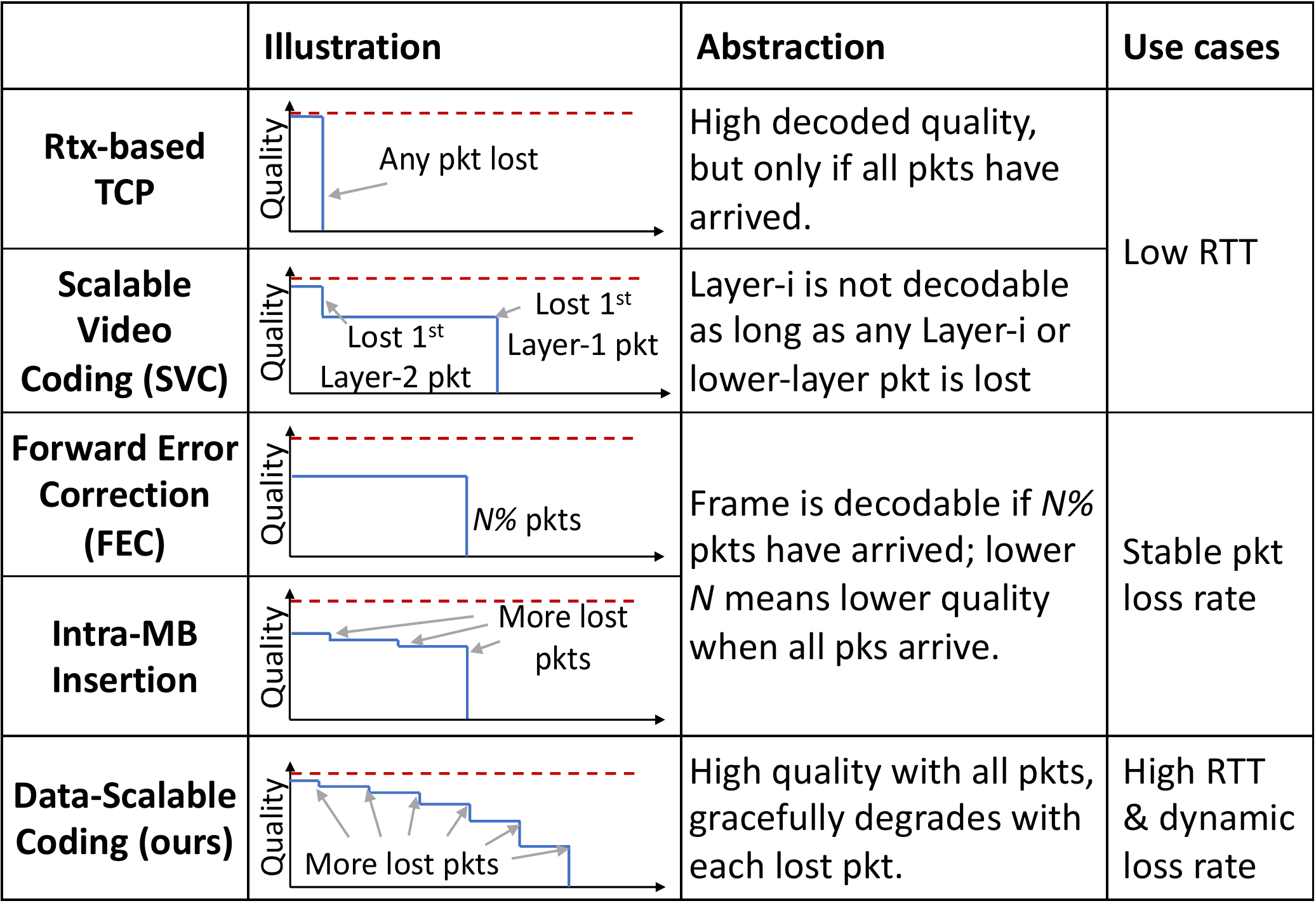}
    \tightcaption{
Summary of loss-resilient schemes. 
    Retransmission-based schemes (TCP, SVC) obtain high quality (the red horizontal line) only if a predefined {\em set} of packets are received.
    Redundancy-based ones (FEC, intra-MB) tolerate a pre-defined maximum number of packet losses by adding redundancy (thus lowering quality when all packets are received).
    In contrast, data-scalable delivery decodes a frame from any non-empty set of packets, improves quality with {\em each} new packet, and has high quality when all packets arrive. 
%    The differences are: 
%    (1) unlike redundancy-based schemes, data-scalable delivery does not need to add more redundancy to tolerate more predicted packet losses, and
%    (2) unlike retransmission-based schemes which must receive a predefined minimum {\em set} of packets, data-scalable delivery can decode a frame with any non-empty set of packets.
%    Comparing different abstractions of loss-resilient schemes. The differences are: 
%    (1) unlike redundancy-based schemes, data-scalable coding does not depend on prediction of packet loss rate; and 
%    (2) unlike retransmission-based schemes which must receive a predefined minimum {\em set} of packets, data-scalable delivery can decode a frame with any non-empty set of packets, and the quality gradually improves with every new packet.
%    (3) unlike redundancy-based ones which must receive a minimum number of packets, data-scalable delivery can decode a frame with 
    }
    \label{fig:primitive-table}
\end{figure}

Loss resilience has long been an important research topic. 
While the details vary greatly among different loss-resilient techniques, it helps to view them through the lens of the high-level abstractions (summarized in Figure~\ref{fig:primitive-table}) that characterize how packets (coded data) are encoded and how {\em decoded video quality of a frame changes with received data (packets)}.
These abstractions also influence the interaction with the remaining parts of a real-time video application, such as congestion control and bitrate adaptation.

\begin{comment}
they have followed some high-level abstractions that characterize how packets (coded data) are encoded and how {\em decoded video quality of a frame changes with received data (packets)}.
%\vspace{-0.05cm}
%\begin{align}
%\texttt{\bf Encode}(\texttt{frame}, \texttt{bitrate}_{target}) & \rightarrow \{\texttt{pkts}\}_{sent}\\
%\texttt{\bf Decode}(\{\texttt{pkts}\}_{received}) & \rightarrow \texttt{quality}
%\label{eq:abstraction}
%\vspace{-0.05cm}
%\end{align}
%of certain loss-resilient performance in terms of video quality as a function of received data (packets). 
Viewing them through the lens of abstractions helps us understand their high-level loss-resilience behavior and  with the remaining of a real-time video application, such as congestion control and bitrate adaptation.

\end{comment}

\myparashort{Retransmission-based}clients
%Packet retransmission is a conventional loss-resilient scheme. 
encode a video frame by a traditional codec (\eg H.264) with a target bitrate and must resend lost packets until a {\em pre-defined set} of packets (pre-determined during video coding) is reliably received. 
%This abstraction can be written as:
%\vspace{-0.05cm}
%\begin{align}
%\texttt{\bf Encode}(\texttt{frame}, \texttt{bitrate}_{target}) &\rightarrow \{\texttt{pkts}\}_{sent}\nonumber\\
%\texttt{\bf Decode}(\{\texttt{pkts}\}_{received}\supseteq\{\texttt{pkts}\}_{predefined}) &\rightarrow \texttt{quality}\nonumber
%\vspace{-0.05cm}
%\end{align}
There are two typical retransmission-based schemes.

\smallskip
{\em Basic TCP-based delivery} must retransmit every lost packet for a frame to be decodable. 
Once all packets are received, the quality will be equal to the original encoded quality.
%This RTX's abstraction can be viewed as ``a frame is non-decodable until all its packets are received, at which point the quality is as high as the original encoded quality.''
%if a few packets are resent between two clients whose RTT is higher than inter-frame interval. 

\smallskip
{\em Scalable video coding (SVC)} encodes a video at multiple quality layers such that a quality layer can be decoded as long as packets of this layer and all lower layers are received. %\footnote{\jc{a note on why we dont consider the redundancies in svc}}
%if a packet of one layer is lost while all packets of lower layers are received, the lower layers can still be decoded.
%Thus, the abstraction of SVC is that ``quality reaches that of the layer $i$ if all packets belonging $i$ and all lower layers have been received.''
However, one single lost packet of a lower layer prevents all higher layers from being decoded. 

With retransmission, video codecs can focus on coding efficiency (\ie maximal quality under a given target bitrate). But it also suffers high tail delays when re-sending a packet takes longer than the jitter buffer\footnote{Usually around 50 milliseconds in WebRTC} and frame interval, causing a visually perceivable lag or a skipped frame.
Even if the base (lowest) layer is padded with redundancy (\eg FEC, explained shortly)~\cite{wang2000error}, any packet loss can still be retransmitted to reach higher quality layers. 
%since a pre-determined set of packets will be always reliably delivered

%However, without any redundancy-based protection, a frame is not decodable unless a pre-defined set of packets (all packets or those of some layers) are received in its entirety. 
%%any lost packets might block the decoding of a frame, making retransmission inevitable in many cases. 
%Thus, retransmission is often inevitable, and retransmission can be very slow in long-RTT networks, especially when the extra delay to request a lost packet is longer than the jitter buffer (\fillme), which result in a visually perceivable lag or skipped frame.
%Therefore, retransmission is also complemented by some redundancy-based loss-resilient techniques, so only fewer lost packets need to be resent. 

\myparashort{Redundancy-based}schemes are used in complement to retransmission-based techniques. 
%are often also complemented with some redundancy-based loss-resilient techniques, so only fewer lost packets need to be resent. 
With this technique, a frame is decodable when at least $N\%$ packets are received. However, the decoded quality is lower with a smaller $N$.
Here, $(1-N\%)$ represents the redundancy rate.
Redundancy-based schemes essentially trade coding efficiency for loss resilience. 
%The abstraction is:
%\vspace{-0.05cm}
%\begin{align}
%\texttt{\bf Encode}(\texttt{frame}, \texttt{bitrate}_{target}, N\%) &\rightarrow \{\texttt{pkts}\}_{sent}\nonumber\\
%\texttt{\bf Decode}(|\{\texttt{pkts}\}_{received}| \geq N\%,N) &\rightarrow \texttt{quality}\nonumber
%\nonumber
%\vspace{-0.05cm}
%\end{align}
Redundancy-based schemes have been studied intensively.
% in the IEEE multimedia community.
%Two typical redundancy-based schemes are FEC and intra-MB insertion.

%This equation entails the efficiency-resilience tradeoff: higher redundancy \texttt{redun} leads to lower $\alpha_{\texttt{redun}}$ (\ie higher loss resilience), but it also means lower resulting quality if $|\{\texttt{pkts}\}_{received}| \geq \alpha_{\texttt{redun}}$ (\ie lower coding efficiency).

%The abstraction of redundancy-based loss resilience is characterized by:
%\vspace{-0.05cm}
%\begin{align}
%\texttt{\bf Encode}(\texttt{frame}, \texttt{bitrate}_{target}, \texttt{redun}) &\rightarrow \{\texttt{pkts}\}_{sent}\nonumber\\
%\texttt{\bf Decode}(|\{\texttt{pkts}\}_{received}| \geq \alpha_{\texttt{redun}},\texttt{redun}) &\rightarrow \texttt{quality}\nonumber
%\nonumber
%\vspace{-0.05cm}
%\end{align}
%This equation entails the efficiency-resilience tradeoff: higher redundancy \texttt{redun} leads to lower $\alpha_{\texttt{redun}}$ (\ie higher loss resilience), but it also means lower resulting quality if $|\{\texttt{pkts}\}_{received}| \geq \alpha_{\texttt{redun}}$ (\ie lower coding efficiency).
%Here, we discuss two concrete redundancy-based schemes.

\smallskip
{\em Forward error correction (FEC)} and Reed-Soloman and Fountain Codes~\cite{mackay2005fountain} are often used in real-time applications.\footnote{FEC is not in the standard of WebRTC. Our evaluation uses the FEC logic implemented in Google's open-source repository~\cite{WEBRTCSRC}.
% \jc{a footnote on the suboptimal use of fec in webrtc}
} %(see \cite{??,??} for an in-depth discussion).
Ideally, these FEC schemes take an encoded video/frame and code it with redundancy bits, and the receiver can recover the data if {\em any} packets contain the same amount of bytes as the original data 
% This is aligned with existing applications, \fillme
that was received. 
%, with \fillme and \fillme achieving near-ideal performance of FEC and fountain codes. 
%Therefore, FEC's abstraction can be described as ``the quality of a frame is low (undecodable) until {\em any} $M$ packets are received.''

FEC is known to have the ``cliff effect''. 
%(a frame or chunk encoded by a video coder like H.265), 
No matter how much redundancy is added, the decoder has to receive at least as many bytes as in the encoded video file, \ie retransmission cannot be avoided if the encoding video bitrate already exceeds the network capacity.
Moreover, the decoded video quality is pre-decided by the video encoder and remains the same even if more data arrives.
Thus, with too little redundancy, the receiver must wait for more packets to arrive or be retransmitted; with too much redundancy, the video quality will be low.
Thus, {\em they are most effective, {\em if} the sender knows how many bytes can be timely delivered in advance}.

%Thus, the video encoder should ideally encode each video frame at exact the bitrate that matches the amount of data received when the frame is decoded. 

%These issues affect Reed-Soloman and fountain codes: that they cannot decode a frame only when a minimum number of packets are received, and the quality will not increase with more arrived packets. 

\begin{figure}[t!]
    \centering
      \begin{subfigure}[b]{0.6\linewidth}
         \centering
         \includegraphics[width=1.1\linewidth]{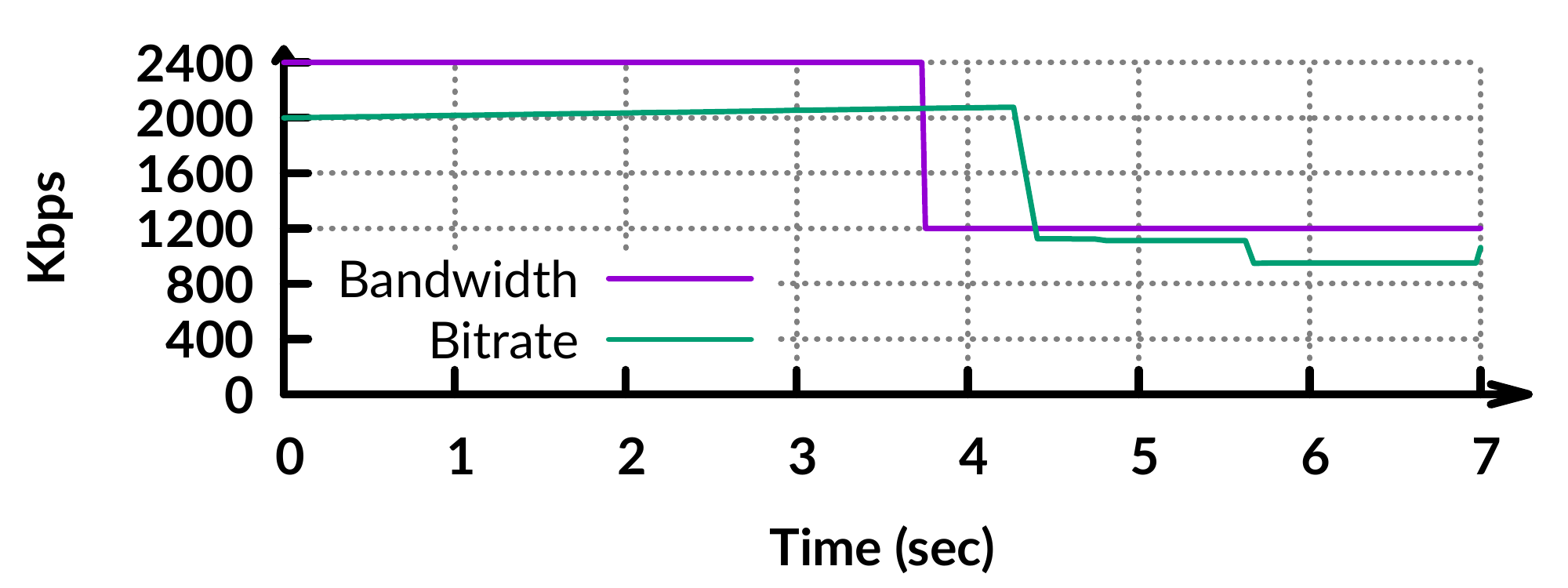}
         \caption{Bandwidth and bitrate}
         \label{fig:}
     \end{subfigure}
     \begin{subfigure}[b]{0.6\linewidth}
         \centering
         \includegraphics[width=1.1\linewidth]{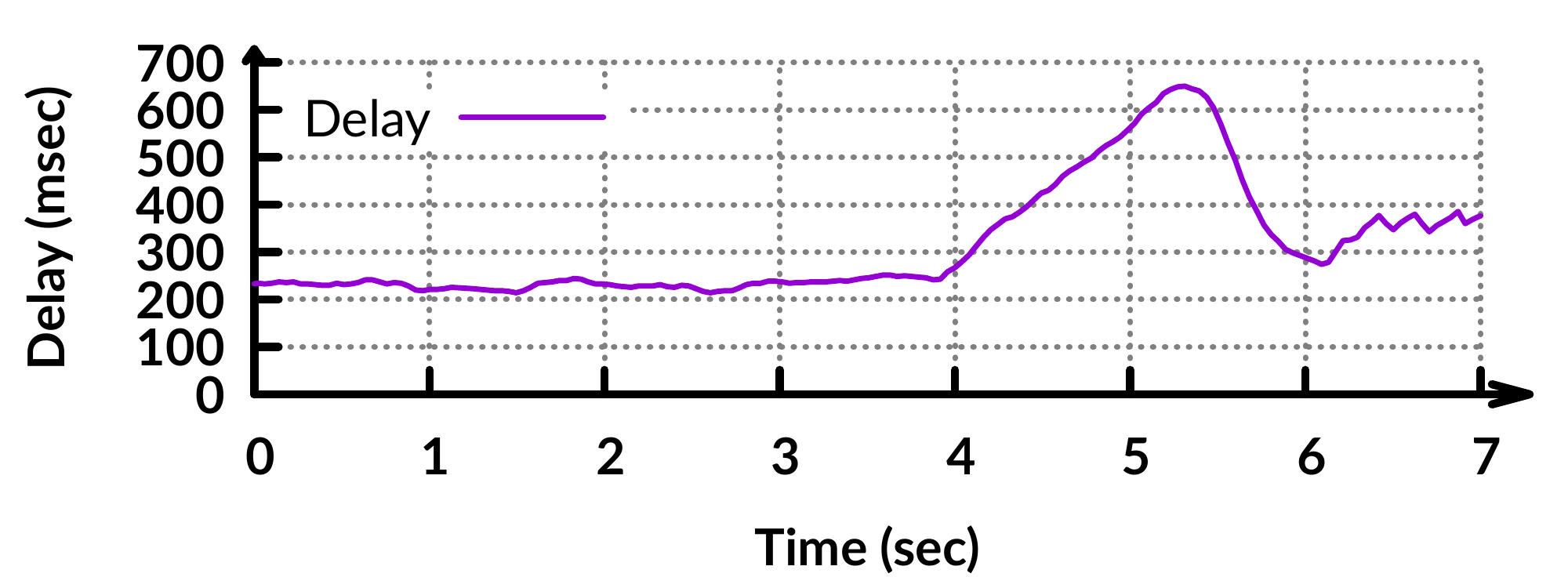}
         \caption{Frame delay}
         \label{fig:}
     \end{subfigure}
    \tightcaption{An WebRTC run with a sudden bandwidth drop}
    \label{fig:motivate-example}
    % \vspace{-0.5cm}
\end{figure}

Unfortunately, it is not always possible to predict the number of missing packets before the receiver decodes a frame. 
Figure~\ref{fig:motivate-example} gives an example showing why setting the FEC redundancy rate is difficult. 
We run WebRTC against a simple bandwidth trace over a network link of RTT 300ms and a drop-tail queue of 1024 packets.\footnote{By default, bandwidth probing and FEC are enabled by WebRTC to allow fast sending adaptation, and the FEC redundancy rate is dynamically determined by the packet loss in the last a few seconds. \cite{holmer2013handling} offer a comprehensive description of WebRTC's congestion control and FEC logic.
}
%Figure~\ref{??} shows that the frame delays (the time between sending a frame and enough bytes received to decode it) is littered with sporadic periods of sudden spikes.
When bandwidth suddenly drops, the sending rate and the FEC redundancy rate adjust only after any sign of congestion is returned to the sender (after at least one RTT), causing many packets to be dropped and delaying subsequent frames. 
The long delays could have been mitigated, if FEC adapts more conservatively (\eg increase redundancy faster but decrease it more slowly). But this will cause lower video quality as more bytes are used for redundancy when bandwidth is stable.
Techniques like dynamic reference selection~\cite{salsify} may mitigate this issue, but could not have avoided the congestion and delay of the first few frames. %as elaborated in \S\ref{??}.
This corroborates the findings in the literature (\eg~\cite{google-doc-fec})

Source-coding-based schemes (\eg intra-MB insertion) do not explicitly add redundancy like FEC. 
% has also proposed techniques to make video coding (\eg H.264) resilient to packet losses by keep more existing pixel-level redundancy during the video coding process (as opposed to FEC which adds redundancy on top of an encoded frame).
Instead, it reduces the inter-dependencies between macroblocks (MBs) that are potentially similar to each other. 
For instance, intra-MB insertion adaptively sets more MBs in \texttt{INTRA} mode, making these MBs and others that refer to them decodable, even if other data is lost.\footnote{Packet losses can also be tolerated by choosing a better reference frame to allow the receiver to skip undecodable frames without affecting subsequent frames, such as in Salsify~\cite{salsify}. These techniques can be used jointly with \name to select better quality reference frames.}
%For instance, \jc{Yihua, explain R/D optimization, intra-MB insertion, dynamic reference selection, etc}
%\jc{need a footnote here to describe how entropy encoding can be stopped at the boundary between packets. }
Both intra-MB insertion and FEC must lower coding efficiency: FEC adds redundancy data and intra-MB insertion keeps more inter-macroblock redundancy.
%Figure~\ref{fig:motivate-regime} illustrates when existing schemes are effective.

In short, retransmission-based schemes are sufficient if the RTT between the clients is low enough to conceal the retransmission delay. 
%compared to the desirable frame delay, packet retransmission would be sufficient to conceal any packet losses.
Redundancy-based schemes are suitable for connections with largely predictable packet loss rates.
% of the next frame is largely predictable, FEC-based schemes will be sufficient. 
However, both of them fall short under high RTT and highly dynamic packet jitters and losses.

\tightsubsection{A new abstraction: Data-scalable delivery}

%To overcome the limitations of prior loss-resilient schemes, 
This work envisions a new abstraction of loss resilience, {\em data-scalable} delivery, and presents a concrete implementation (\S\ref{sec:training}).
Data-scalable delivery should meet two criteria.
{\em (i)} First, the frame is decodable once {\em any} non-empty subset of its packets are received, and quality improves with more packets received (or equivalently, the quality degrades gracefully with more packets missing).
{\em (ii)} Second, unlike the redundancy-based approach, how a frame is encoded should not depend on the prediction of future packet losses.
%{\em (ii)} Second, the decoded quality with {\em any} $x\%$ ($x>0$) received packets is lower but still comparable to classic codec (\eg H.265) with $x\%$ FEC redundancy.
%Note that classic coder $x\%$ with  can be viewed as a reference design that optimizes compression efficiency, not loss resilience.
%\item 
%\end{packeditemize}
At a high level, this new abstraction can be expressed as follows:
\vspace{-0.05cm}
\begin{align}
\texttt{\bf Encode}(\texttt{frame}, \texttt{bitrate}_{target}) &\rightarrow \{\texttt{pkts}\}_{sent}\nonumber\\
\texttt{\bf Decode}(|\{\texttt{pkts}\}_{received}|) &\rightarrow \texttt{quality}\nonumber
\nonumber
\vspace{-0.05cm}
\end{align}

\myparaq{What's new}
Figure~\ref{fig:primitive-table} contrasts data-scalable delivery with other loss-resilient techniques.
\begin{packeditemize}
\item
Unlike FEC (RS or Fountain Code), data-scalable delivery allows the receiver to decode after receiving any packet and obtain higher quality with each new packet received.

\item
While SVC might increase quality with more received packets, it is not data-scalable, because it imposes a hierarchical structure of packets. 
One missing packet may render multiple quality layers undecodable (even if the base layer is protected by FEC~\cite{SVCFEC1, schierl2007using}).
%Each new packet does not always improve quality, until all packets of a layer and the lowers layers have been received. 
%As we will show in \S\ref{sec:eval}, SVC (even with FEC-protection on the base layer as commonly used~\cite{??,??}) does not always improve quality with each new packet, until all packets of a higher layer (and all layers below it) have been received. 
%In other words, a missing packet may render multiple quality layers undecodable. 

\end{packeditemize}

We should stress that any loss resilience comes at the expense of coding efficiency, and {\em data-scalable delivery is no exception}.
We expect that the quality in absence of any packet losses will be slightly lower than traditional codecs.

%,  illustrates an example frame whose coded data is delivered in six packets arriving at the receiver at an arbitrary order. 

%It helps to highlight the difference between data-scalable delivery and SVC whose quality also increases with more packets. 
%However, SVC is not exactly data-scalable, because it imposes a strict hierarchy among packets.
%As we will show in \S\ref{sec:eval}, SVC (even with FEC-protection on the base layer as commonly used~\cite{??,??}) does not always improve quality with each new packet, until all packets of a higher layer (and all layers below it) have been received. 
%In other words, a missing packet may render multiple quality layers undecodable. 

%For this reason, in practice, SVC often employs extra loss-protection like FEC but only to protect its base layer (known as unequal error protection).

%The most noticeable benefit of data-scalable delivery is that, 
%Therefore, while the exact arrival time of each packet is unknown to the receiver, 
%Data-scalable delivery offers the receiver the {\em flexibility} to decode a frame anytime after receiving at least one packet of the frame, and the quality always improves with each new packet.

%We should clarify that this work does {\em not} invent a new loss tolerant paradigm. 

\tightsubsection{Why autoencoder?}
\label{subsec:why-autoencoder}

Our approach to data-scalable delivery is based on autoencoders, an emergent class of neural-network coders for video (I-frames and P-frames).
The recent autoencoders borrow ideas from traditional codecs (\eg compressing motion vectors and residuals separately, rather than compressing each frame from scratch), and replace handcrafted heuristics with multi-layer neural nets (NNs) which are trained on large sets of videos (\eg Vimeo-90K~\cite{vimeo-dataset}). 
Recent autoencoders' performance has significantly improved~\cite{dvc, elf}. 
%Thanks to these trends, their compression efficiency and generalization have been improving over the recent years.

%Recent autoencoders have improved their compression efficiency, thanks to higher-capacity encoder/decoder DNN architectures, which can be trained on large set of videos (\eg Vimeo-90K~\cite{??}). 
%
%%Like traditional codecs, recent video autoencoders compute and compress motion vector estimation and residuals separately, rather than compressing each frame from scratch. 
%These optimizations improve not only compression efficiency but also generalization of these NN-based autoencoders.

\begin{figure}[t!]
    \centering
%    \begin{subfigure}[b]{0.49\linewidth}
%         \centering
%         \includegraphics[width=\linewidth]{figs/psnr-loss-vs-pretrained.pdf}
%         \caption{P-frame model}
%         \label{fig:}
%     \end{subfigure}
%     \hfill
%     \begin{subfigure}[b]{0.49\linewidth}
%         \centering
%         \includegraphics[width=\linewidth]{figs/autoencoder_pretrained_qmap_v1.png}
%         \caption{I-frame model}
%         \label{fig:}
%     \end{subfigure}

         \includegraphics[width=0.5\linewidth]{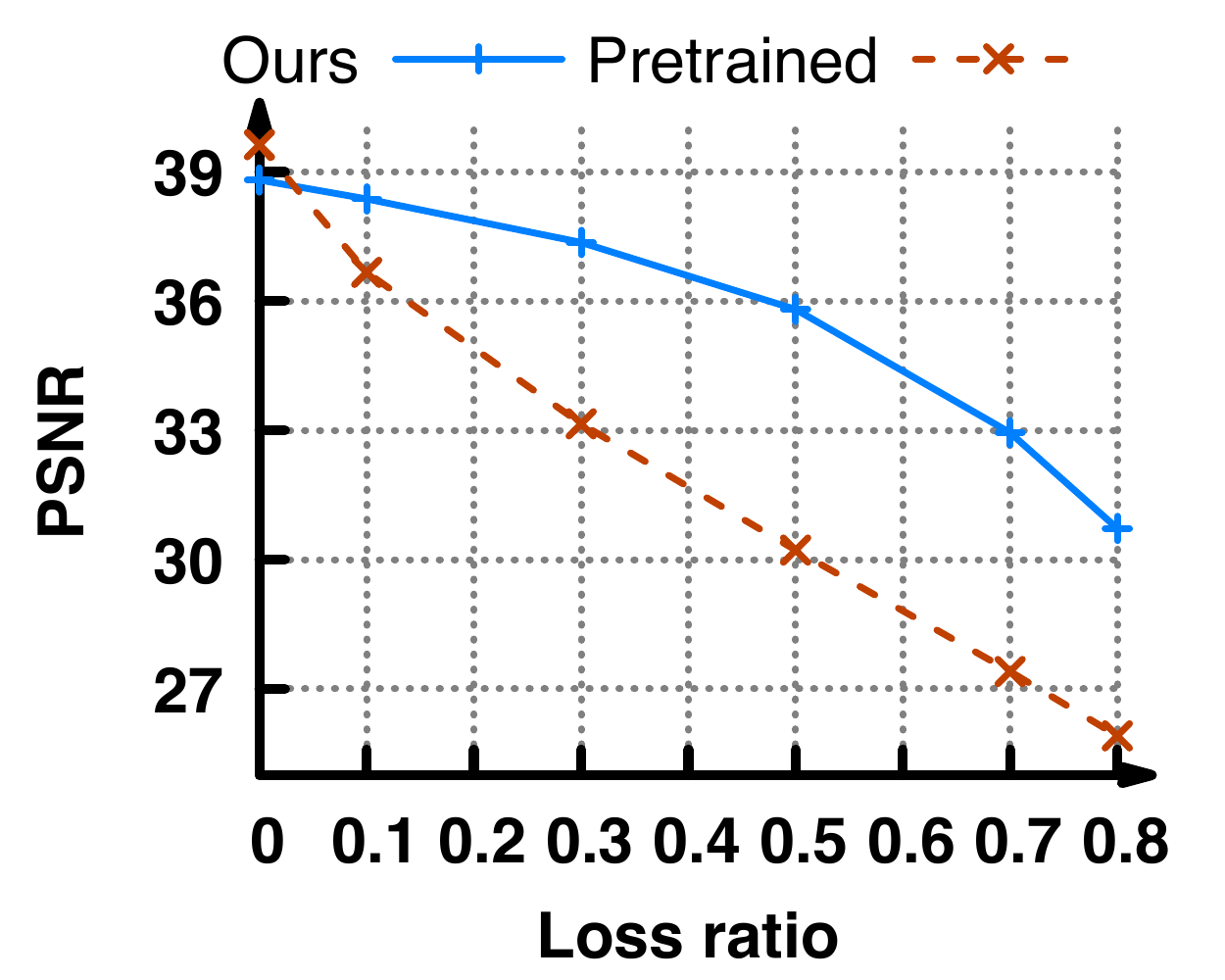}

    \tightcaption{Loss resilience of an example pre-trained autoencoder (red). It is already data-scalable (decodable with any subset of packets), but the quality drops much faster than the graceful degradation observed in our solution. 
    % \jc{YIHUA, make the figure wider. Swap the order of the legend to "Pre-trained autoencoder" and "Our custom autoencoder"} \jc{also add a horizontal line of H.264/265 without loss}
%    Our models has a more graceful quality degradation under packet losses comparing to pretrained models
    }
    \label{fig:autoencoder-pretrained}
\end{figure}

%\mypara{Autoencoder-based data-scalable coding}
We choose to use autoencoders not because of their coding efficiency (which comes with caveats discussed shortly), but because their quality under data erasure (albeit suboptimal) {\em might} fit the definition of data-scalable delivery. 
Figure~\ref{fig:autoencoder-pretrained} shows an example: we use a pre-trained P-frame autoencoder~\cite{dvc} to encode frames from a video (not from the training set) and test its decoded quality with an increasing rate of random data erasure (we will explain how this simulates packet losses in \S\ref{subsec:sim-losses}). 
%To emulate the effect of different packet loss rates, we randomly zero \fillme-\fillme\% of values in the autoencoder-returned data (before entropy encoding). 
The red curve shows that as more data is received, the decoded quality gradually improves. 
%(To avoid overfitting, we test it on a different video dataset than the training dataset.)
%It is very encouraging that even a pre-trained autoencoder leads to a (suboptimal) data-scalable delivery.

However, as the figure shows, the pre-trained autoencoder degrades quite sharply with more data erasure. 
(After all, these autoencoders are not trained to perform well under random data erasure.) 
In \S\ref{sec:training}, we will show how to train data-scalable autoencoders that produce better quality under various packet losses (\eg the blue line in Figure~\ref{fig:autoencoder-pretrained}).
%Our goal in the remaining of the paper is two-fold: (1) creating better data-scalable autoencoders that produce better quality under various packet losses (see the blue line in Figure~\ref{fig:autoencoder-pretrained}), and (2) designing a custom real-time video delivery system that uses autoencoders as a codec, to better balance between frame delay and video quality. 

\tightsubsection{Caveats of using autoencoders}

Despite being emergent and not as heavily optimized as traditional codecs, autoencoders have gained increasing attention in the signal processing community for image compression~\cite{bourtsoulatze2019deep,gunduz2019machine} and wireless communication~\cite{erpek2020deep,dai2020deep}, by the multimedia community for high-volumetric video compression~\cite{le2022mobilecodec,li20213d}, and recently by the networking community for chunk-based video streaming~\cite{swift}, though these efforts still rely on traditional schemes (like FEC) to handle packet losses. 

Though autoencoders are studied extensively, there has been much less attention given to a thorough comparison between autoencoders and traditional codecs (H.265 and H.264, which have been extremely optimized in speed and efficiency).
This could lead to potentially biased (and sometimes exaggerated) perceptions of autoencoders' performance. 
%Traditional video codecs have been extremely heavily engineered, but 
For instance, if one tests an autoencoder and H.265 under the same small (4-10) GoP, autoencoders will outperform H.265 because small GoPs align better with how autoencoders are trained but not with H.265.
These comparisons could also be unfair to autoencoders as well.
For instance, H.26x searches optimal reference MBs from {\em multiple} frames to reduce residuals, whereas most autoencoders derive motion vectors and residuals from a {\em single} reference frame (which is not a fundamental constraint) and instead, they focus on compression of given motion vectors and residuals.
Moreover, autoencoders can be trained to optimize either PSNR or SSIM, so papers evaluate only the metric that the model is trained for.
%whereas traditional codecs are designed with YUV-SSIM in mind\cite{some citation?}.

This work reports our efforts to make autoencoders loss resilient (\S\ref{sec:training},\ref{sec:delivery}) as well as practical (in speed and memory usage) for real-time videos (\S\ref{subsec:optimize}).
In our evaluation of autoencoders, we have tried our best to avoid pitfalls (\eg those listed in~\cite{codec-eval-pitfalls}). (For space limitation, we elaborate the details in \S\ref{subsec:eval:setup}.)
For instance, we report both PSNR and SSIM and use H.26x's default encoding setting without limiting GoP, and stress-test autoencoders on video sets different from the training set.
We also use a common setting of real-time video~\cite{ZEROLAT1,ZEROLAT2} with the \texttt{fast} preset in H.265 and no B-frames or rc-lookahead.
%disable B-frames (0 bframe and rc-lookahead) and use \texttt{fast} preset in H.265.
% \jc{yihua: please find some citation/pointers to justify this choice.}
%train autoencoders on the same dataset (Vimeo-90K~\cite{??}) but report the test result on different datasets.
%Moreover, we allow H.26x can still use multiple history frames as reference, but only one frame as reference in autoencoders. 
%To be clear, we make sure that this is still unfavorable to autoencoders, since H.26x can still use multiple history frames as reference, but autoencoders only use one.
%The bottom line is that in our testing, our pre-trained autoencoder~\cite{??,??} (which \name customizes) it is on par with H.265 and better than H.264, and we have also confirmed that in the most aggressive encoding mode (\eg with B-frames and \texttt{fast}), H.265 achieves higher compression efficiency than autoencoders. 

We hope to present a balanced view in this work: even though the autoencoders have lower coding efficiency than the heavily engineered video codecs like H.265, we believe that their potential to reduce tail delay (via data-scalable delivery) outweighs the slightly lowered quality, which is also being constantly improved by better neural-network models.

\newcommand{\Coder}{\ensuremath{C_\phi}\xspace}
\newcommand{\Decoder}{\ensuremath{D_\theta}\xspace}
\newcommand{\Quality}{\ensuremath{F}\xspace}
\newcommand{\Packetize}{\ensuremath{A}\xspace}
\newcommand{\X}{\ensuremath{\mathbb{X}}\xspace}
\newcommand{\x}{\ensuremath{\textbf{x}}\xspace}
\newcommand{\y}{\ensuremath{\textbf{y}}\xspace}
\newcommand{\z}{\ensuremath{\textbf{z}}\xspace}
\newcommand{\loss}{\ensuremath{P_{\textrm{loss}}}\xspace}

\tightsection{\newae: Data-Scalable Autoencoder}
\label{sec:training}

%\name is a concrete instantiation of data-scalable delivery. 
At a high level, \name uses a new data-scalable autoencoder, \newae and a custom delivery framework that leverages the new autoencoder (\S\ref{sec:delivery}) to balance tail delay and quality.
%We will describe the autoencoder design in this section, followed by the delivery framework in the next section. 

\tightsubsection{Basic formulation of \newae training}

%We begin with the preliminary of autoencoders. 
We denote the encoder (a neural network parameterized by weights $\phi$) by $\Coder$ and the decoder (a neural network parameterized by weights $\theta$) by $\Decoder$.
For an input frame $\x$ (a $c\times w\times h$ tensor), the coded data $\z=\Coder(\x)$ will be sent to the receiver, which then runs the decoder to reconstruct the original frame from $\z$: $\hat{\x}=\Decoder(\z)$.
Note that the encoder and decoder performing the lossy reconstruction are {\em separate} from the lossless entropy coding or FEC coding. (We will discuss entropy encoding and packetization in \S\ref{subsec:sim-losses}.)
We retrain the DVC autoencoder~\cite{dvc} to encode P-frames, and discuss how to add I-frames in \S\ref{subsec:delivery:iframe}.
(We do not consider B-frames as they are not commonly used in low-latency videos.)
%We customize two existing autoencoder architectures, \cite{dvc} for P-frames (with motion vectors and residuals as input) and \cite{qmap} for I-frames (with a static image as input).\footnote{We do not consider B-frames are they are not commonly used in low-latency videos.} 
%We borrow their neural network architectures (their architectures in details in \S\ref{app:arch}).
%Since they share the same conceptual framework described above, we do not discuss them separately unless necessary.

\mypara{Conventional autoencoder training}
An autoencoder is trained to minimize the ``loss'' function (not ``packet loss'') $\Quality(\hat{\x},\x)= Distortion(\hat{\x},\x)+\alpha\cdot Size(\hat{\x})$, which is a weighted sum of its pixel distortion $Distortion(\hat{\x},\x)$ (L2-norm of $\hat{\x}-\x$)\footnote{Pixel-wise distortion may not be the best way to capture human perception of reconstructed images, but it is often used in autoencoder training for its differentiability. We will evaluate the trained autoencoders in terms of more standard metrics, such as SSIM and PSNR (in RGB and YUV).} and its entropy-encoded size $Size(\hat{x})$ (\eg estimated by a pre-trained neural net).
A smaller $\alpha$ means that the output of $\Coder(\x)$ will tend to have better visual quality in $\Decoder(\z)$ but also higher bitrate. 
(We present \name's bitrate adaptation in \S\ref{subsec:optimize}.)
%To adapt to the dynamic sending rate, multiple versions of the autoencoders are trained, each with a different weight $\alpha$, and they share most (backbone) layers, which significantly reduces the memory footprint when switching between bitrates. (See \name's bitrate adaptation in \S\ref{subsec:optimize}.) 
% A higher $\Quality$ value is better (higher visual quality and/or lower entropy-coded size).
Since $\Coder$, $\Decoder$, and $\Quality$ are differentiable, the encoder and decoder NNs can be trained as an end-to-end architecture on a training video set $\X$:
\vspace{-0.08cm}
\begin{align}
\textrm{\bf max~~~~} \mathbb{E}_{\x\sim \X} \Quality(\Decoder(\z), \x)\textrm{, where }\z = \Coder(\x)
\label{eq:1}
\vspace{-0.08cm}
\end{align}

%\jc{anton/yihua, can you draw a simplified figure to depict the input/output of the used autoencoders? the figure should also include the tensor dimensions of the input and output. }

%\begin{figure*}[t!]
%    \centering
%         \centering
%        \includegraphics[width=0.95\linewidth]{figs/per-packet-encoding.pdf}
%    \caption{\name's packetization and per-packet entropy (arithmetic) encoding. When a packet is lost, the elements mapped to it will be set to zeros.}
%    \label{fig:per-frame}
%\end{figure*}

\mypara{Training data-scalable autoencoder}
The training process in Eq.~\ref{eq:1} can be seen as having no packet loss, \ie the decoder's input equals the encoder's output. 
To improve coding efficiency (low reconstruction distortion with fewer bits) under various packet loss rates, we use a custom training process that uses a random distribution $\loss$ to simulate packet losses:
\vspace{-0.1cm}
\begin{align}
\textrm{\bf max~~~~} \mathbb{E}_{\x\sim \X} \Quality(\Decoder(\z), \x)\textrm{, where }\color{purple}
\underbrace{\highlight{red}{$\z\sim \loss(\z | \Coder(\x))$}}_{\text{\sf \footnotesize \textcolor{purple!85}{Simulating packet losses}}}
\label{eq:2}
\vspace{-0.1cm}
\end{align}
%where the change is highlighted in color.
$\loss$ takes the encoder output $\Coder(\x)$ and returns the distribution of the data seen by the decoder after packet erasures under some \rate. 
%As we will see next, with appropriate randomization, $\loss$ can simulate the effect of packet losses on data received by the decoder.
Simulating packet losses in Eq.~\ref{eq:2} raises three unique questions, and we will address them next.
\begin{packedenumerate}
\item How should $\loss$ convert $\Coder(\x)$ to data packets to simulate the effect of dropping a packet? 
%This decides what data seen by the decoder would be if packets are randomly dropped. 
\item Given that the function $\loss$ may not be differentiable, how to train autoencoders with Eq.~\ref{eq:2}? 
\item What packet loss rates should be simulated in training so that the trained model can handle unseen loss rates?
\end{packedenumerate}
%It is important to note that these questions are not involved in canonical autoencoder training, but they become critical here for the need to simulate packet losses in the training process.
% \section{Simulating \rates in autoencoder training}

% \mypara{How to simulate packet losses}

%\mypara{Simulating packet losses in autoencoder training}
%Training a data-scalable autoencoder using Eq.~\ref{eq:2} faces three key questions:
%\begin{packedenumerate}
%\item How to make Eq.~\ref{eq:2} differentiable for a wide range of $\loss$ that simulates various packet loss patterns?
%\item How to convert $\Coder(\x)$ to data packets? This decides how losing a packet would affect the data seen by the decoder.
%\item What \rates should be simulated in training, so that the trained model can deal with a range of unseen \rates?
%\end{packedenumerate}
% The first question concerns packetization of the coder output and the second question concerns what \rates should be used to train the model that can deal with a range of \rates.

% , so that the trained model can handle packet losses of various \rates in real networks. 

\begin{figure}[t!]
    \centering
         \centering
        \includegraphics[width=0.99\linewidth]{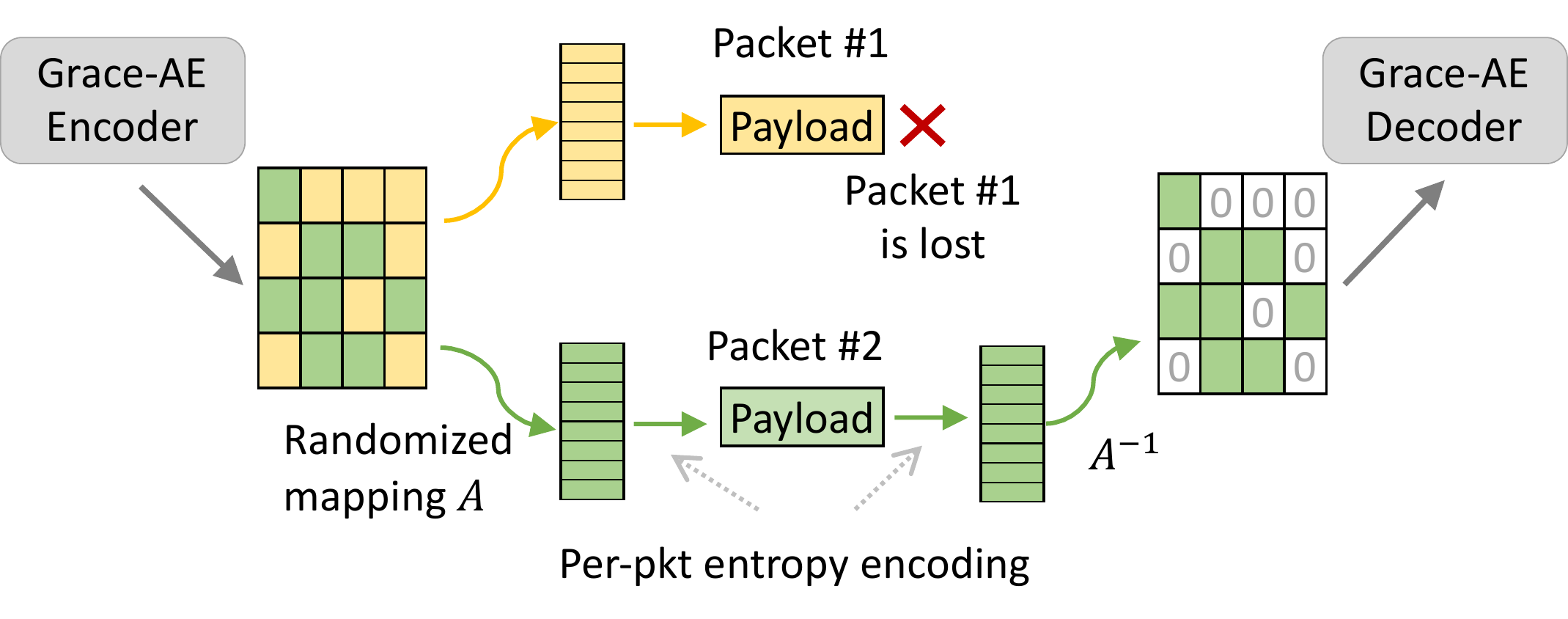}
    \tightcaption{\name's packetization and per-packet entropy (arithmetic) encoding. When a packet is lost, the elements mapped to it will be set to zeros.}
    \label{fig:per-frame}
\end{figure}

\subsection{\hspace{-0.2cm}Simulating packet losses during training}
\label{subsec:sim-losses}

\mypara{Packetization and data erasure}
First, to packetize the encoder output $\z=\Coder(\x)$, \name uses a {\em packetization function} $\Packetize$ (illustrated in Figure~\ref{fig:per-frame}) to map each element position in $\z=\Coder(\x)$ (a 3-D or 4-D tensor) to one of $k$ non-overlapping lists\footnote{\name uses 8, 16, 24 depends on the size of the input frame} and then losslessly encodes the elements in each list to a bitstream of a packet.
Conversely, depacketization first losslessly decodes each packet to a list of elements and maps elements back to their original positions in $\z$ by the inverse function $\Packetize^{-1}$.
One can use an off-the-shelf entropy coder to losslessly code the elements in a packet to a bitstream. 
With {\em per-packet} entropy coding, each packet is individually decodable. 
Per-packet entropy encoding is commonly used in intra-MB insertion (explained in \S\ref{subsec:abstractions}) and other source-coding loss resilient schemes. 
We use arithmetic coding~\cite{torchac} to encode each packet. 
%Because the elements in each packet are selected randomly from the entropy encoder's input, 
Because the elements in each packet share the same distribution as all elements in $\z$ (as we will show shortly), different packets share the same arithmetic coder (without needing a separate arithmetic distribution per packet), so each list has the same number of elements, the encoded packet size of the list will be roughly the same.

%This is different from typical autoencoders which entropy encodes the whole $\z$ tensor before packetization and thus any lost packet will make the received data not decodable by the entropy decoder. Moreover, since we use arithmetic coding (\fillme), so the total size of individually encoded packets is similar to the output of coding $\z$, if the elements in each packet share the same distribution as all elements in $\z$ (which is fortuitously ensured by random mapping. 

%(We discuss in \S\ref{app:entropy-encoding} \jc{need an appendix here} how to fit the bitstream in the normal packet size of 1.4KB.)

If a packet is lost, \name will set each element whose position is mapped to the lost packet to zero. 
Thus the packetization function $\Packetize$ decides how missing one packet would affect the coded data seen by the receiver.  
\name's packetization function is a {\em uniformly random} mapping (A pseudo-random function which has a reversible mapping\footnote{A reversible mapping allows the decoder to map elements from packets (including lost packets) back to the vector $\z$. An example can be mapping the $i^{\textrm{th}}$ element to the $j=(i*p\textrm{ mod }n)^{\textrm{th}}$ packet at the $(i*p-j)/n^{\textrm{th}}$ position, where $n$ is the number of packets and $p$ is a prime number.}). 
As a result, with a 30\% packet loss rate, $\loss(\z|\Coder(\x))$ will produce a distribution of $\z$ that randomly zeros 30\% of elements in $\z$.

\begin{figure}[t!]
    \centering
     \begin{subfigure}[b]{0.49\linewidth}
         \centering
         \includegraphics[width=\linewidth]{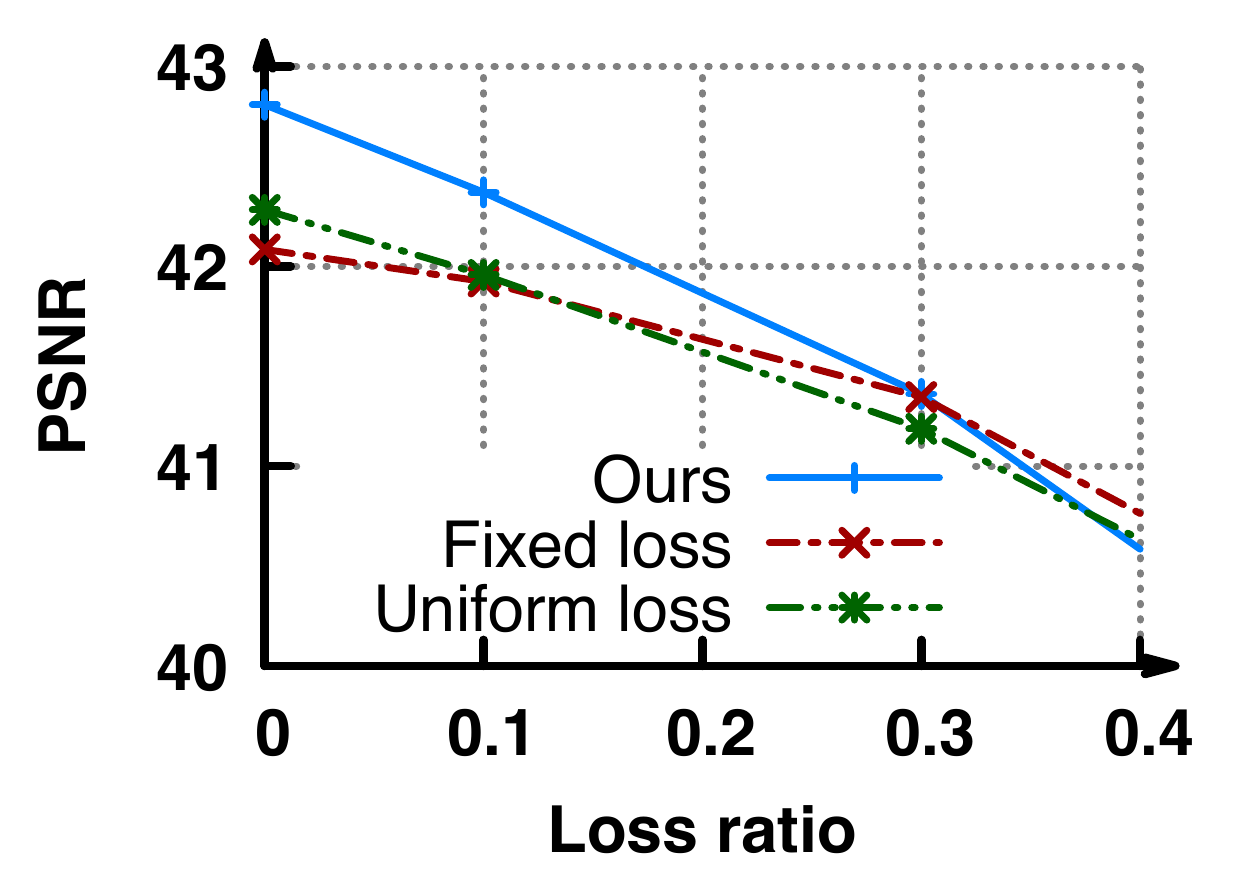}
         \caption{Impact of different data loss distribution used during training}
         \label{fig:training_loss_distribution}
     \end{subfigure}
     \hfill
     \begin{subfigure}[b]{0.49\linewidth}
         \centering
         \includegraphics[width=\linewidth]{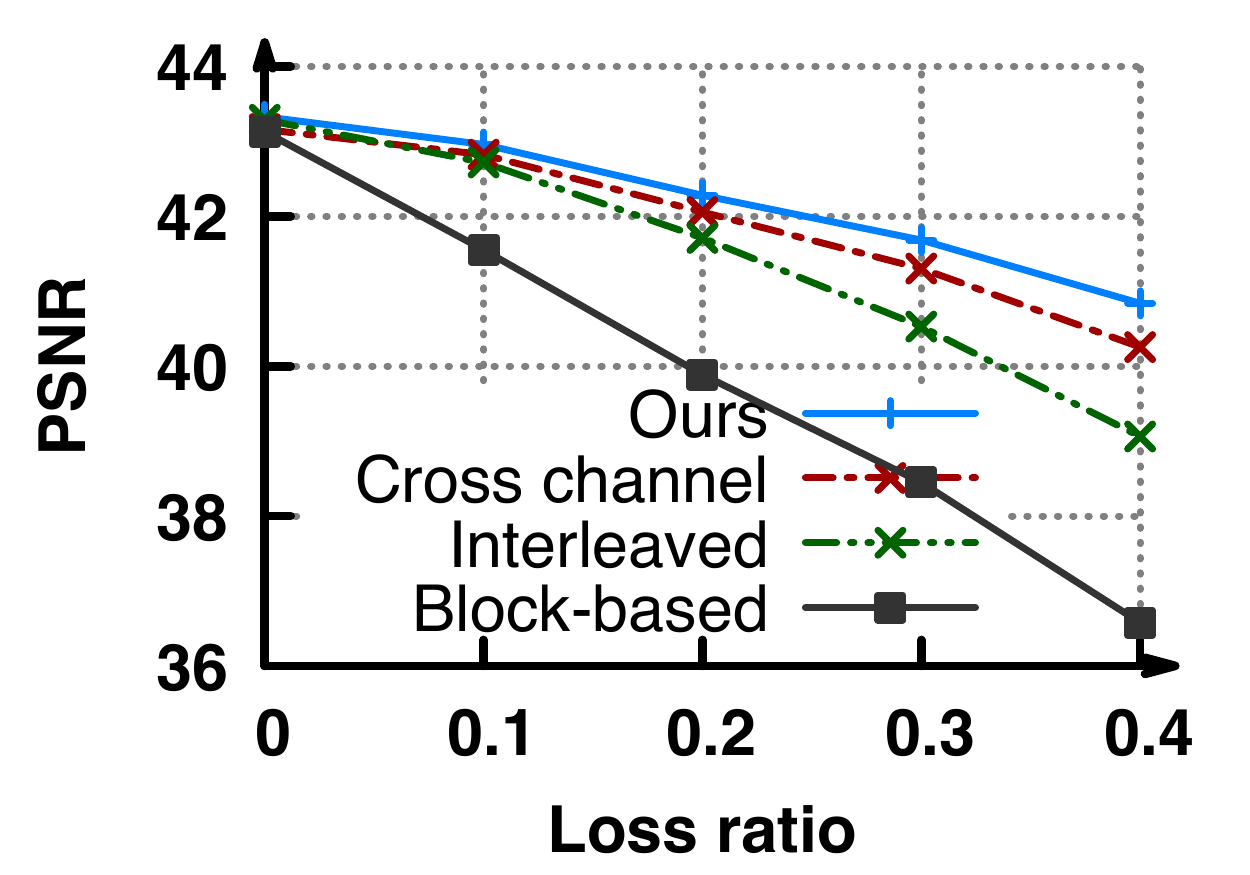}
         \caption{Impact of different data loss patterns used during training}
         \label{fig:training_loss_pattern}
     \end{subfigure}
     \vspace{.05cm}
     \tightcaption{Empirically comparing different ways of simulating packet losses.}
    \label{fig:training_loss_simulation}
\end{figure}

\mypara{Why random packetization}
The choice of randomized packetization empirically works better than alternatives, for two reasons. 
\begin{packeditemize}
\item In an autoencoder output $\z$, some elements have more impact on the decoded quality (defined in \S\ref{subsec:understanding}).
%, and zeroing each important element has disproportionally higher impact on the decoded quality than zeroing other elements.
A random mapping thus makes sure that losing $x\%$ packets will affect $x\%$ important elements (whereas non-randomized packetization may affect more than $x\%$ important elements).
\item A random mapping ensures that all elements in $\z$ have the same distribution as those mapped to each packet. This makes it easier to maintain roughly the same entropy-encoded size of each packet. 
\end{packeditemize}

Figure~\ref{fig:training_loss_distribution}(a) empirically compares \newae trained with the random packetization and \newae trained with alternative schemes, such as block-based packetization (\ie each packet contains a contiguous block of elements in $\z$) and interleaving-based packetization. 
Both are suboptimal, since important elements can be next to each other and have a fixed distance from each other under $\z$'s specific tensor structure, so in the worst case both block-based packetization and interleaving packetization can drop a higher fraction of important elements than packet loss rate.

\mypara{Choice of packet loss rate in training}
At first glance, we should simulate as many packet loss rates during training as possible, so that the model can deal with various packet losses.
However, our empirical results (Figure~\ref{fig:training_loss_simulation}(b)) suggest that adding too much randomness in \loss might make the training converge to suboptimal autoencoder models. 
Specifically, training \newae with just a few values of packet loss rates generalizes better than with a wide range of packet loss rates, even on unseen test packet loss rates.
Moreover, the packet loss rates in training should include 0\% (otherwise the trained autoencoder will perform significantly worse than the pre-trained autoencoder on low packet loss rates.

\mypara{Making \newae trainable}
Since $\loss$ is a non-differentiable random function, the gradient of the expectation of $\Quality$ in Eq.~\ref{eq:2} cannot be directly calculated. 
To address this issue, we use the REINFORCE trick~\cite{Kingma2014} for reparameterization. 
We express the gradient of Eq.~\ref{eq:2} as
% According to differentiation property of logarithms, we can 

$$ \nabla_\phi \loss(\z) =  \loss(\z) \nabla_\phi \log{\loss(\z)}$$

where $\loss(\z)$ is our packet loss distribution so our gradient of the expectation of $\Quality$ becomes
\begin{multline}
\nabla_\phi \mathbb{E}_{x\sim \loss(\z)}([\Quality(\Decoder(\z), \x)]) \\ 
= \mathbb{E}_{\z\sim \loss(\z)}([\Quality(\Decoder(\z), \x)\nabla_\phi \log{\loss(\z)}])
\end{multline}
which can be estimated using Monte-Carlo sampling
$\approx \frac{1}{N} \sum_{i=1}^{N} \Quality(\Decoder(\z_i), \x) \nabla_\phi \log{\loss(\z_i)} $
Since in our application the loss is independent and identically distributed random variable, the gradient evaluates to either $0$ or $1$, hence we propagate the gradients for the encoder only for $\Quality(\Decoder(\z_i), \x)$ where $\loss(\z_i) = 1$.

\tightsubsection{Why is \newae data-scalable?}
\label{subsec:understanding}

%As illustrated in Figure~\ref{fig:autoencoder-pretrained}, compared to the pre-trained autoencoder (which is already data-scalable), \newae (re-trained on the same video dataset but with packet losses) achieves much higher quality on unseen videos under various test packet loss rates, even those that are {\em not} seen during training. 
%This remarkable improvement begs the question why \newae can be much more resilient to packet losses. 

We believe that the reasons are two-fold. 
First, compared to SVC, FEC, or H.26x-based coding, the elements in \newae's encoding output (before entropy encoding) do not impose any strict structure (\eg hierarchy or dependencies). 
Instead, the relationship between elements in \newae output is ``flat'', and changing any element (to zero) would not render the frame non-decodable. 

\newae's decoder can be seen as a special case of variational autoencoders (VAE)~\cite{pu2016variational}, which are generative neural networks trained to create realistic images from a {\em random point} in the input space. 
In the case of \newae, our random point follows the distribution of $\loss$. 

Second, compared to the pre-trained autoencoder, individual elements in \newae's output have a lower impact on the decoding quality, so quality degrades more gracefully with more lost data.
Though the NNs are too complex to show this directly, we can indirectly illustrate it as follows.
For a frame, we first calculate the ``gradient'' of decoding quality with respect to each element in $\z$.
%Mathematically, the gradient of decoding quality with respect to elements in $\z$ is a vector:
$gradient=\frac{\partial \Quality(\Decoder(\z),\x)}{\partial\z}$.
Then, we sort the elements in descending order of their gradients. 
% If top $k$ elements are zeroed, the incremental impact on decoding quality is marginal. 
As more top $k$ elements are zeroed, we observe that the quality of the pre-trained autoencoder drops sharply, whereas that of \newae drops much more slowly (not shown for space limitation).

%\jc{
%\begin{itemize}
%    \item \newae's performance is surprising, but can be explained.
%    \item flat relationship among elements, no strict hierarchy/dependencies like in svc or fec
%    \item saliency is more robust to losses: figure~\ref{fig:per-frame}
%\end{itemize}
%}

\tightsection{Real-time video delivery over \newae}
\label{sec:delivery}

\name is built on the data-scalable autoencoder \newae. 
%\name is a concrete real-time video communication system built on a data-scalable autoencoder (\newae).
We focus on unique issues arising from the use of \newae as the codec, including how to deliver each frame to strike a desirable delay-quality tradeoff, and how to deliver a sequence of frames when congestion or packet losses affect multiple frames. 
The choice of the congestion control logic is complementary to \newae. 
In \S\ref{subsec:eval:e2e}, we will evaluate \name with different congestion control logic, such as GCC in WebRTC~\cite{carlucci2016analysis} or Salsify CC~\cite{salsify}. 
We do not claim that \name's strategies are new (\eg reference frame synchronization has been used in~\cite{salsify}). 
Nonetheless, \name integrates concrete design choices driven by the need, in order to unleash the full potential of \newae and its data-scalable codec primitive.

\tightsubsection{Encoding a sequence of frames}
\label{subsec:delivery:iframe}

%To stream a sequence of video frames, a key question is how often I-frames should be inserted in P-frames. 
Placement of I-frames poses a particular challenge for \name's autoencoders (both pre-trained and re-trained), for two reasons: (1) I-frames are generally bigger in size than P-frames, and 
(2)  while the first few P-frames after an I-frame enjoy high quality, quality gradually degrades after about 10 frames. 
%(As explained in \S\ref{subsec:ae-background}, \name uses only I-/P-frames, as in most real-time video systems.)
%For the P-frame autoencoder we have tried, we observe that the decoded quality can be very high right after each I-frame but it gradually degrades after about 10 frames. 

A strawman would add I-frames frequently (say every 10 frames)\footnote{Many computer-vision papers on autoencoders have conveniently used a small GoP when compared against H.265, but the implication of large I-frame size for congestion control is rarely discussed.}.
Even though this might achieve a decent tradeoff between quality and {\em average} bitrate (comparable to H.264 and H.265 with large I-frame intervals), sending I-frames (which have bigger sizes than the P-frames in between) frequently will make it difficult for congestion control to send the frames at a fixed interval (inversely proportional to the frame rate).

\begin{figure}[t!]
    \centering
    
        \includegraphics[width=0.85\linewidth]{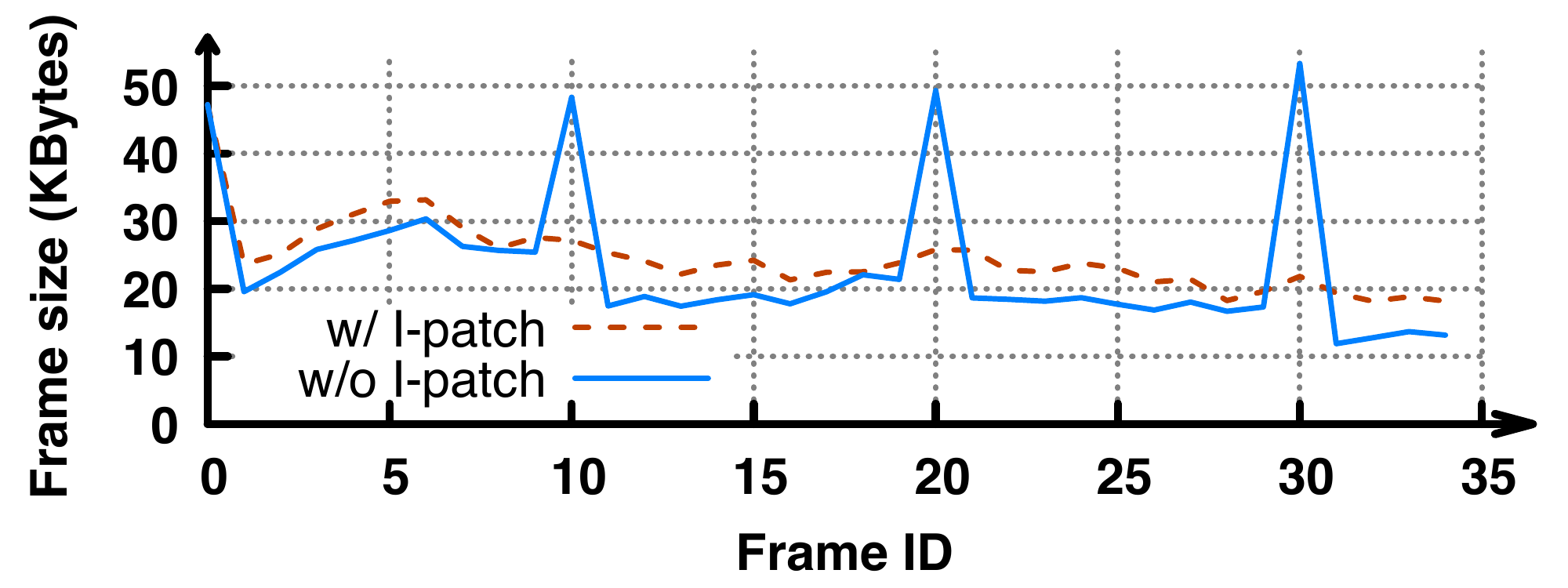}
    \tightcaption{Encoding each P-frame with a small I-patch leads to smoother frame sizes than naively inserting I-frames.}
    \label{fig:keyframe_smooth}
\end{figure}

To add I-frames frequently without the intermittent bitrate spikes caused by the I-frames, \name encodes a small patch (about 128$\times$128 to 512$\times$512) as a tiny I-frame, which we call {\em I-patch}, on {\em every frame} %(the pixels outside the I-patch will be encoded as a P-frame)
(an I-patch will not impact the original P-frame), and the location of the I-patch on a frame changes over time such that the I-patch will scan through the whole frame size every $k$ frames. 
Depending on the original frame size, $k$ varies between 6 and 20. 
This scheme effectively ``amortizes'' the size of I-frames across $k$ frames, thus smoothing the frame size as well as quality across frames (illustrated in Figure~\ref{fig:keyframe_smooth}).
Moreover, since the I-frame autoencoder's encoding delay is proportional to the input size, encoding a small I-patch adds only a small compute overhead for each frame.\footnote{We should clarify that design choice of frequent I-frames works well for autoencoders, but {\em not} for classic codecs for two reasons. First, an I-frame generated by our autoencoder is only 2-5$\times$ bigger than a P-frame, whereas this ratio in H.264 is 4-10$\times$. Second, the P-frame quality in H.264 is relatively stable, whereas the quality of a P-frame is high following an I-frame (or an I-patch).}

\tightsubsection{Packet losses across multiple frames}

An important benefit of adding an I-patch (a tiny patch encoded as an I-frame without referring to previous frames) at a different place in each frame is: each patch will be encoded as an I-patch every $k$ frames. 
This has a significant implication for loss resilience over multiple consecutive frames. 

%\mypara{Encoding/decoding synchronization}
An acute reader may realize a potential problem arising from encoding and decoding being asynchronous. 
For instance, when the sender encodes the $10^{\textrm{th}}$ frame as a P-frame, it needs the decoded the $9^{\textrm{th}}$ frame as the reference (to compute motion vectors and residuals), and without knowing which of the $9^{\textrm{th}}$ frame's packets will be lost, it will assume that the $9^{\textrm{th}}$ frame is decoded with all packets being received. (This is common to classic codecs and autoencoders.)
However, if the receiver decodes the $9^{\textrm{th}}$ frame with only a subset of its packets, the actual reference frame used to decode the $10^{\textrm{th}}$ will differ from what the encoder has assumed. 
Thus, most codecs require all packets of the $9^{\textrm{th}}$ frame to be received before decoding it (thus longer delay), or let the sender encode the $10^{\textrm{th}}$ based on an earlier reference (say the $5^{\textrm{th}}$) whose lost packets are already known to the sender. 
The latter {\em synchronizes} the encoding and decoding states (like in Salsify~\cite{salsify} during packet losses), but it also inflates the frame size, since the difference between the $10^{\textrm{th}}$ and the $5^{\textrm{th}}$ frames  is much greater than between the $10^{\textrm{th}}$ and the $9^{\textrm{th}}$ frames.

%the P-frame encoding of the $10^{\textrm{th}}$ frame will have a low quality when being decoded with a different reference frame than the encoder has assumed. 

\name uses a simple yet efficient solution---it does not always synchronize the encoding/decoding state on every single frame; instead, it relies on the frequent I-patches (explained in \S\ref{subsec:delivery:iframe}) to synchronize the encoding/decoding states in each patch-size region. 
This works for two reasons.
First, without synchronizing the encoding and decoding states, \newae's quality degradation increases only marginally when multiple consecutive frames (less than about 10) are affected by packet losses.
For instance, applying a 50\% packet loss rate on one single frame can reduce PSNR by 3-4, and applying the same loss rate on 5 consecutive frames reduces PSNR by 7-9. 
%\jc{this corroborates our empirical improvements in \S\ref{??}}
%quality degrades by \fillme\% when 10 consecutive frames are affected by 50\% packet losses without the clients synchronizing their encoding/decoding states. 
Second, each patch will be encoded as an I-patch every $k$ frames ($k$ is between 6 and 20), so the impact of packet losses will not last for more than $k$ frames if the next I-patch is not lost. 
That said, in the worst case, if the I-patches of the same region are dropped persistently, this could cause bad decoded pixel values (\ie a small patch of white pixels), though we have observed it extremely rarely.

\begin{figure}[t!]
    \centering
         \centering
        \includegraphics[width=0.995\linewidth]{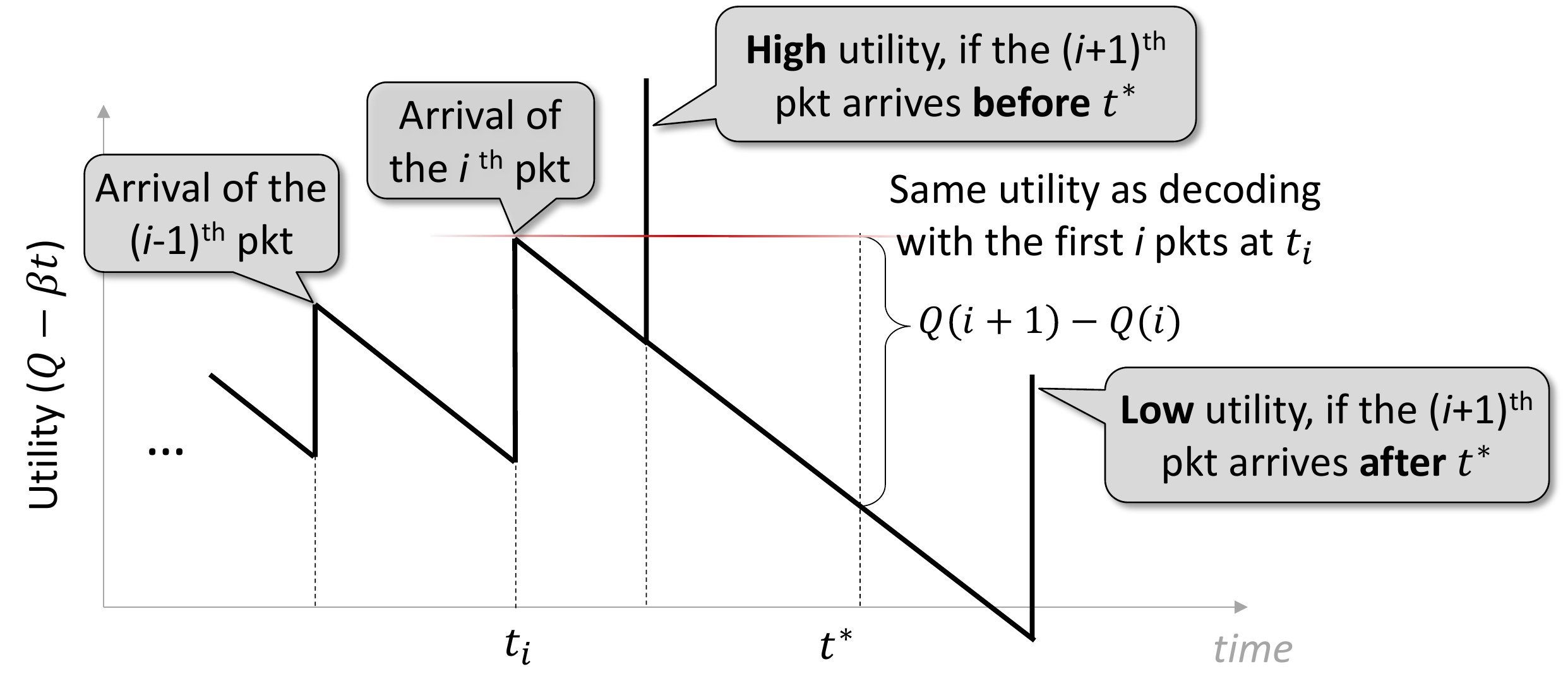}
    \tightcaption{
    \name waits for the next frame until a deadline $t^*$, after which the utility is lower even if the next packet arrives.
    % At the arrival of the $i^{\textrm{th}}$ packet, \name decides the maximum wait time for the next packet.
    }
    \label{fig:decode-time}
\end{figure}

\tightsubsection{Choosing when to decode a frame}

The data-scalable abstraction allows the receiver to flexibly decide, as packets arrive at the receiver, whether to decode the frame using currently arrived packets or wait for more packets in hope of improving the quality. 
(In contrast, with FEC, the receiver will always wait until enough data is received.)
%(In other loss-resilient schemes (\S\ref{??}), the receiver must wait for more packets until the data is decodable.) 
%\name embraces this flexibility with the following logic to trade frame delay with decoded quality. 

To make this decision, \name defines the {\em utility} of a frame as a linear combination of decoding quality and frame delay (since the natural decoding time) on each frame: $Q(|\{\textrm{pkts before }t\}|)-\beta\cdot t$, where $t$ is when the frame is decoded, $Q(n)$ is the frame's decoding quality with $n$ packets 
The precise values of $Q(n)$ are frame-dependent, but they are roughly similar across frames, so \name estimates $Q(n)$ using history frames.
%, and $\tau$ is the frame's natural decoding time (\eg 40ms after the last frame was decoded with frame rate 25fps).
%In our experiment, we set parameter $\beta=\fillme$.
This objective works well in practice, though there might be other objectives that better capture users' quality perception. An optimal design of such utility is beyond the scope of this paper.

With this utility in mind, Figure~\ref{fig:decode-time} illustrates \name's logic: on receiving the $i^{\textrm{th}}$ packets of a frame (not necessarily the $i^{\textrm{th}}$ sent packet) at time $t_i$, the receiver will immediately decode the frame if all packets of the frame have arrived; otherwise, it will wait for the next packet of the frame until a deadline $t^*=t_i+\frac{1}{\beta}\cdot(Q(i+1)-Q(i))$, and if it has not received the $(i+1)^{\textrm{th}}$ packet by then, it will decode the frame with $i$ packets. 
The reason is that if the next packet comes at time $t_{i+1}$ after $t^*$, the utility $Q(i+1)-\beta\cdot (t_{i+1}-\tau)$ will be lower than the utility $Q(i)-\beta\cdot (t_i-\tau)$ if the receiver does not wait at all\footnote{This is because {\scriptsize $Q(i+1)-\beta\cdot (t_{i+1}-\tau)<Q(i+1)-\beta\cdot (t^*-\tau)=Q(i)-\beta\cdot (t_i-\tau)$}}. 
%Here, $\beta$ is a tunable parameter, and $Q(n)$ is the quality of the frame based on $n$ packets.
This strategy makes intuitive sense: the receiver will wait longer if the improvement in quality by getting one more packet (\ie $Q(n+1)-Q(n)$) is higher, but it will not wait for too long.

\tightsubsection{WebRTC integration}
\name is implemented with 3K lines of code, in both Python (mostly for autoencoder NNs) and C++ (for frame delivery and WebRTC integration). The code and trained model of \name will be made public upon the publication of this paper.
The integration with WebRTC is logically straightforward since \newae (including I-frame and P-frame encodings) exposes similar interface as the default codec in WebRTC. 
%\jc{Ziyi, please fill in the details}

We substitute the libvpx VP8 Encoder/Decoder in WebRTC with our \newae implementation. When the sender encodes a frame, it parses the image data from the \texttt{VideoFrame} data structure (YUV format) into \texttt{torch.Tensor} (RGB format) and feed it into our \newae encoder, which will return the encoded result as a byte array. Then the encoded bytes are stored into an \texttt{EncodedImage} (class in WebRTC) and sent through the network to the receiver as RTP packets. 
% \zy{I could also add several sentences about packetization logic (commented out below)}
% The encoded bytes are then packetized as RTP packets and sent through the network to the receiver. 
We modify the built-in \texttt{RtpVideoStreamReceiver} (class in WebRTC) so that the receiver could flexibly decode the received packets even when not all the packets are received. When the receiver decides to decode the frame, it depacketizes the received packets into encoded data. Then it will use the \newae decoder to decode the image into RGB format and then convert it back to YUV for displaying on the receiver side.

%\jc{need to explain the gains are not directly from the entropy encoder or motion vector estimation}

\tightsubsection{Optimizations}
\label{subsec:optimize}

While \newae is functionally similar to traditional codecs, it faces a few technical challenges to function as the codec in a real-time video client.
Note that these challenges are not specific to loss resilience, so we do not consider packet losses in the rest of the section.

\mypara{Reducing compute overhead}
Most research on autoencoder focuses on coding efficiency (using fewer bits to get higher quality), but the encoding/decoding speed is much lower than the heavily optimized codecs like H.265.
Fortunately, we show that on a single NVIDIA GeForce 3080 GPU, \name's encoding speed can be greatly improved: from 5~fps on 720p HD (and 11~fps on 480p) videos by the off-the-shelf implementation to 18~fps (and 40~fps). More results in \S\ref{subsec:eval:system}. 
\name achieves such speedup by downsizing the image (by default $4\times$) during motion-vector estimation and motion compensation, and then upsample the motion-compensated image before calculating and encoding the residuals. 
This speeds up encoding because motion-vector estimation and motion compensation are the slowest steps during P-frame encoding and their delays are proportional to the input size.
Thus, the overall encoding delay will be reduced if they are fed with downsized images. 
At the same time, because the residuals are still computed between the full-resolution image and the motion-compensated image, the coding efficiency is not affected by this change for most frames.

\mypara{Bitrate control}
In real-time video, throughput estimation can vary even between two consecutive frames, and the encoded frame size should be barely below the target frame size.
This has been a challenge with traditional codecs.
%Thus, it is important that most frames are smaller than the target size set by congestion control, but traditional codecs are not able to do that. 
%This has been observed in \cite{salsify}, which proposes to encode each frame at multiple quality levels in order to find an encoding that fits the target bitrate. 
\name provides bitrate control logic similar to~\cite{salsify} to encode a frame with a maximum size below a target. 
We train multiple autoencoders, each using a different weight of frame size $\alpha$ in its training loss function (\S\ref{subsec:sim-losses}).
These resulting autoencoders cover a range of bitrate range. 
%\fillme$\times$ difference between the highest bitrate and lowest bitrate. (A wider bitrate range is possible with more autoencoders.)
As long as the target bitrate is within this range, \name should be able to select the autoencoder whose encoded frame size is barely lower than the target. 
A naive solution will run multiple autoencoders (like~\cite{salsify}), but the compute overhead will increase linearly.
Fortunately, we can again re-use the motion-compensated image to get the residuals, and only run different versions of residual encoders to encode the residuals with different bitrates.
Since residual encoder is faster than the motion estimation and compensation, the speed up is significant.

\mypara{Reducing memory footprint}
For the client to switch between multiple autoencoders (with different bitrates), their memory footprint should be minimized, so more of them can be loaded into GPU memory to avoid the delay of loading and unloading models. 
\name trains multiple autoencoders by fine-tuning the last 10-25\% layers of the same model (\ie they share the backbone layers), so only the one autoencoder and the last few layers of different autoencoders need to be loaded in GPU memory. 
This reduces the memory footprint of the models by 60\% compared to when all models are trained separately without sharing any layers.
A similar idea has also been used in Swift~\cite{swift} in multiple SVC layer encoders.

\tightsection{Evaluation}
\label{sec:eval}

The findings of our evaluation can be summarized as follows:
\begin{packeditemize}
\item {\bf Loss resilience:} Though \newae's quality under no packet loss is on par with H.264 and slightly worse than H.265, it is data-scalable---graceful quality degradation with higher loss rate---under various videos and bitrates. 
%\jc{a few example numbers}
\item {\bf Quality-delay tradeoff:} On real and synthetic network traces with bandwidth fluctuation, \name reduces tail delay by 2$\times$, compared to other loss-resilient schemes (FEC, Salsify) while still obtaining decent quality.
%strikes a better trade-off between average video quality and tail delay than a wide range of baselines.
%\jc{a few example numbers}
\item {\bf Overhead:} Our implementation of \name can encode/decode 480p video at 40~~fps/50~fps and 720p HD video at 18~fps/30~fps, a 1.5$\times$ speedup compared to off-the-shelf autoencoders.
\end{packeditemize}

\tightsubsection{Setup}
\label{subsec:eval:setup}

\mypara{Baseline implementations}
Our baseline codecs include H.265 and H.264. 
We use their implementations in \texttt{ffmpeg} (version 4.2.7). 
Importantly, we do not restrict their GoP (I-frame interval) and the number of history frames to search for reference macroblocks. 
Following recent work in real-time video coding~\cite{ZEROLAT1,ZEROLAT2}, we use the \texttt{zerolatency} option (no B-frames) and the \texttt{fast} preset of H.265\footnote{The command line we used to encode a video is \texttt{ffmpeg -y -i Video.y4m -c:v libx265 -preset fast -tune zerolatency -x265-params "crf=Q:keyint=3000" output.mp4} where {\em Q} controls the quality and size of the output video}.

We use implementations for SVC, Salsify, and FEC, based on H.265.
We simplify the implementation in a way that makes their quality slightly better than they would actually do. 
Our implementation of SVC assumes that each quality level exactly {\em matches} the quality and size of the non-SVC coding of H.265, whereas even the state-of-the-art SVC (based also on autoencoders~\cite{swift}) achieves lower quality than H.265 at a quality level of the same frame size. (This effectively is an idealized version of Swift~\cite{swift}.)
We follow Salsify's logic for congestion control, frame skipping, and packet pacing, but instead of using Salsify's multi-trial encoding to approximate the target bitrate of each frame, we assume that the encoded frame size exactly {\em matches} each frame's target bitrate but the quality matches the {\em qp} value whose bitrate is barely {\em higher} than the target bitrate.
For FEC-protected H.26x, we use the same loss-rate prediction logic implemented in the public release of WebRTC~\cite{WEBRTCSRC} but remove the upper-bound of 50\% redundancy rate that is used in WebRTC \eg if the predicted packet loss rate is 60\%, we assume FEC will tolerate any packet loss ratio under 60\%.) %but instead of mapping it to the coarse-grained FEC redundancy rates, we assume an idealized FEC that can reconstruct a frame if the actual loss rate matches the predicted loss rate (\eg if the predicted packet loss rate is 50\%, we assume FEC will exactly tolerate 50\% packet loss.)
%This effectively approximates the situation where data can be packetized infinitesimally.
While these simplifying assumptions lead to a slightly optimistic estimate of the baselines' performance, we hope they make our results more relevant in the long run, as the implementations of SVC, and bitrate control, and FEC might evolve and improve in the future.

\mypara{Network and compute conditions}
We test \name on 8 real bandwidth traces distributed with the Mahimahi network-emulation tool~\cite{netravali2015mahimahi}\footnote{The data set has 16 traces. We filtered out the traces having an average bandwidth lower than 1Mbps as they do not fit the bandwidth requirement of video traffic.}. 
%and we filter traces .
By default, we set the one-way network propagation delay to be 100ms and the queue size to be 25 packets.
We also show the impact of different propagation delays and queue sizes in \S\ref{subsec:eval:e2e}.
We conduct all experiments on a machine with two Nvidia GeForce RTX 2080 GPUs as the sender and receiver (each using one GPU). The server runs Ubuntu 18.04 with 2 Intel Xeon Silver 4110 CPU and 64GB memory.
%\jc{more details, such as cuda version, linux version, cpu, memory, etc}
We measure \name's encoding and decoding delays on the GPU (see results in \S\ref{subsec:eval:system}) and we include these delays in \name's frame delay.

\mypara{Metrics}
We measure the performance of a video session in three aspects.
Quality of a frame is measured by SSIM and PSNR (RGB and YUV), and we report the average value across received frames. 
Delay of a frame is measured by the time lapse between the beginning of encoding to the end of decoding. 
For skipped frames (by Salsify), their decoding time equals the next decoding time of the next received frame. 
We report the tail (95th\% percentile) of the delays across frames in a session.
%\jc{describe frame rate metric}

\mypara{Test videos}
Our evaluation uses 60 videos from three public datasets (Kinetics~\cite{Kinetics}, UVG~\cite{mercat2020uvg} and FVC~\cite{FVCDATASET}), each video is of at least 10 seconds. 
%Table~\ref{sec:??} summarizes the datasets.
Importantly, we choose these videos because they are different from the training set, allowing us to stress-test the generalization of autoencoders.
The FVC dataset includes 5 videos captured from video conferencing, one of \name's main target use-cases.
These videos cover a range of spatial and temporal dynamics (\S\ref{app:siti})
and cover three resolutions (640$\times$360, 1280$\times$720 and 1920$\times$1280), with at least 5 videos in each resolution. 
The performance is reported as an average across all the frames in the videos.

%\mypara{Baselines}
%We compare the performance of \name with multiple state-of-the-art real-time video techniques that use different combinations of video coding and congestion control algorithms: 
%\begin{packeditemize}
%\item {\em Reference:} H.26x with adaptive FEC and GCC
%\item {\em Salsify:}
%\item {\em Idealized SVC:}
%\end{packeditemize}

% \begin{figure}[t!]
%     \centering
%          \centering
%         \includegraphics[width=0.99\linewidth]{figs/psnr-bpp-example.pdf}
%          \label{fig:}
%     \caption{an example. need more of it in different datasets}
%     \label{fig:}
% \end{figure}

% \begin{figure}[t!]
%     \centering
%     \begin{subfigure}[b]{0.49\textwidth}
%          \centering
%          \includegraphics[width=\textwidth]{figs/psnr-vs-loss-360p.pdf}
%          \caption{360P videos (bitrate = 1.4Mbps)}
%          \label{fig:}
%      \end{subfigure}
%      \hfill
%      \begin{subfigure}[b]{0.49\textwidth}
%          \centering
%          \includegraphics[width=\textwidth]{figs/psnr-vs-loss-HD.pdf}
%          \caption{1080P videos (bitrate = 6.2Mbps)}
%          \label{fig:}
%      \end{subfigure}
%      \hfill
%      \caption{\newae's loss resilience on videos with different resolutions}
%     \label{fig:loss_by_resolution}
% \end{figure}

\begin{figure}
    \centering
    \includegraphics[width=\linewidth]{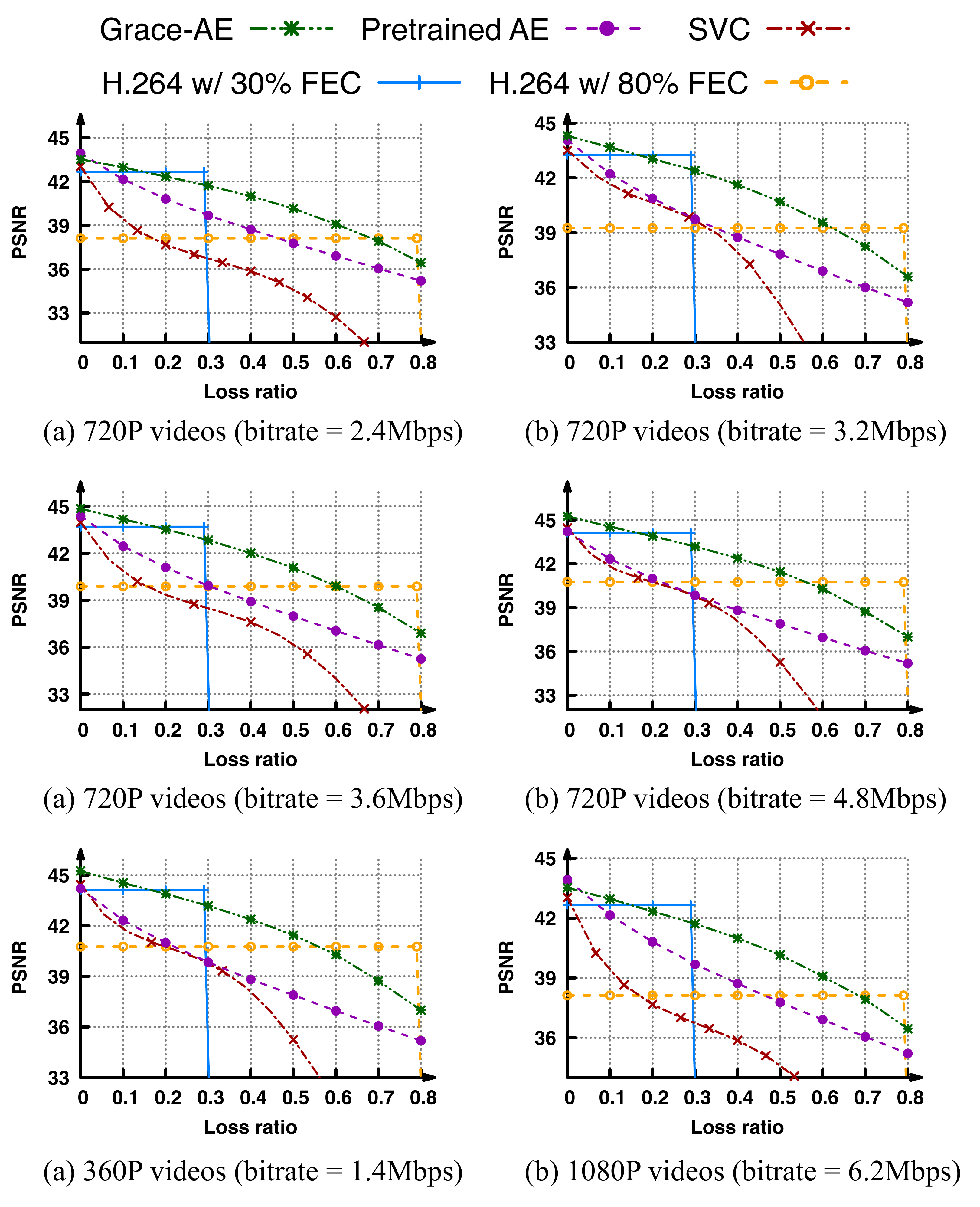}
    \tightcaption{\newae's loss resilience by bitrate and resolution}
    \label{fig:loss_by_resolution_bitrate}
\end{figure}

\tightsubsection{Coding efficiency and loss resilience}

% \name's gains stem from the performance of the autoencoder \newae.
Figure~\ref{fig:loss_by_resolution_bitrate} compares \newae with the baselines codecs\footnote{We do not put Intra-MB insertion baseline here, since it's quality will quick drop below x-axis when loss ratio is between 5\% to 10\%.} under different packet loss rates when encoding videos of the same resolution and same target bitrate. 
In each figure, all lines have almost the same encoding bitrate with difference controlled within 5\%.
We ensure that \newae's encoded bitrate is {\em never} higher than the baselines, so even if the actual encoded bitrate is different, it does not favor \newae.
% Similarly, Figure~\ref{fig:loss_by_bitrate} compares the same set of solution when they have the same encoding bitrate.

% For each figure, the encoded bitrates of different solutions are controlled within 10\% difference (and to make it less favorable to \newae, we make sure that \newae's encoded bitrate is never higher than the baselines).
% Figure~\ref{fig:loss_by_resolution} shows the results by video resolution and Figure~\ref{fig:loss_by_bitrate} shows the results by encoding bitrate on 720P videos\footnote{We do not put Intra-MB insertion baseline here, since it's quality will quick drop below x-axis when loss ratio is between 5\% to 10\%}. 
The most salient feature of \newae is the graceful quality degradation with more packet losses. 
The average PSNR drop is only 2.1 to 3.9 under 30\% and 50\% packet losses. Even when the loss ratio is as high as 80\%, it has an average PSNR of 36.4, which is much better than the baselines except H.264 with 80\% FEC.
Compared with H.264 with fixed FEC protection, \newae can have a higher PSNR of 1.1 to 2.0 among different resolutions and different bitrates when there is no loss. This is because any arrived bytes in \newae contribute to the quality while in FEC, the redundant data is useless even if they do arrive at the decoder. On the other hand, FEC-based loss resiliency is also in a dilemma that a low FEC rate cannot protect against the high loss, but a high FEC rate leads to poor quality when the packet loss ratio is small. Our \newae solve's such dilemma by the graceful quality degradation.
\newae is also better than SVC with unequal FEC protection (\S\ref{subsec:abstractions}). 
The quality of SVC decreases quickly when the loss rate is small, as the high-quality layers are unlikely to be decodable in such cases. 
% As shown in Figure~\ref{fig:loss_by_resolution_bitrate}, \newae has a average PSNR gain of 3.9 over SVC when the loss ratio is 30\%.
We also note that the pretrained autoencoder models, which have not seen any losses in training, degrade poorly with higher loss rates. 
% It has an average PSNR decrease of 4.3 when loss ratio is 30\%, which is 2$\times$ higher than \newae.
% to h264 + fec baseline
% comparing to SVC

%\jc{
%\begin{itemize}
%    \item by bitrate
%    \item by video set
%    \item psnr vs. bitrate
%\end{itemize}
%}

% \jc{need a footnote to explain why intra-mb insertion is so bad}

\begin{figure}[t!]
    \centering
         \centering
        \includegraphics[width=0.7\linewidth]{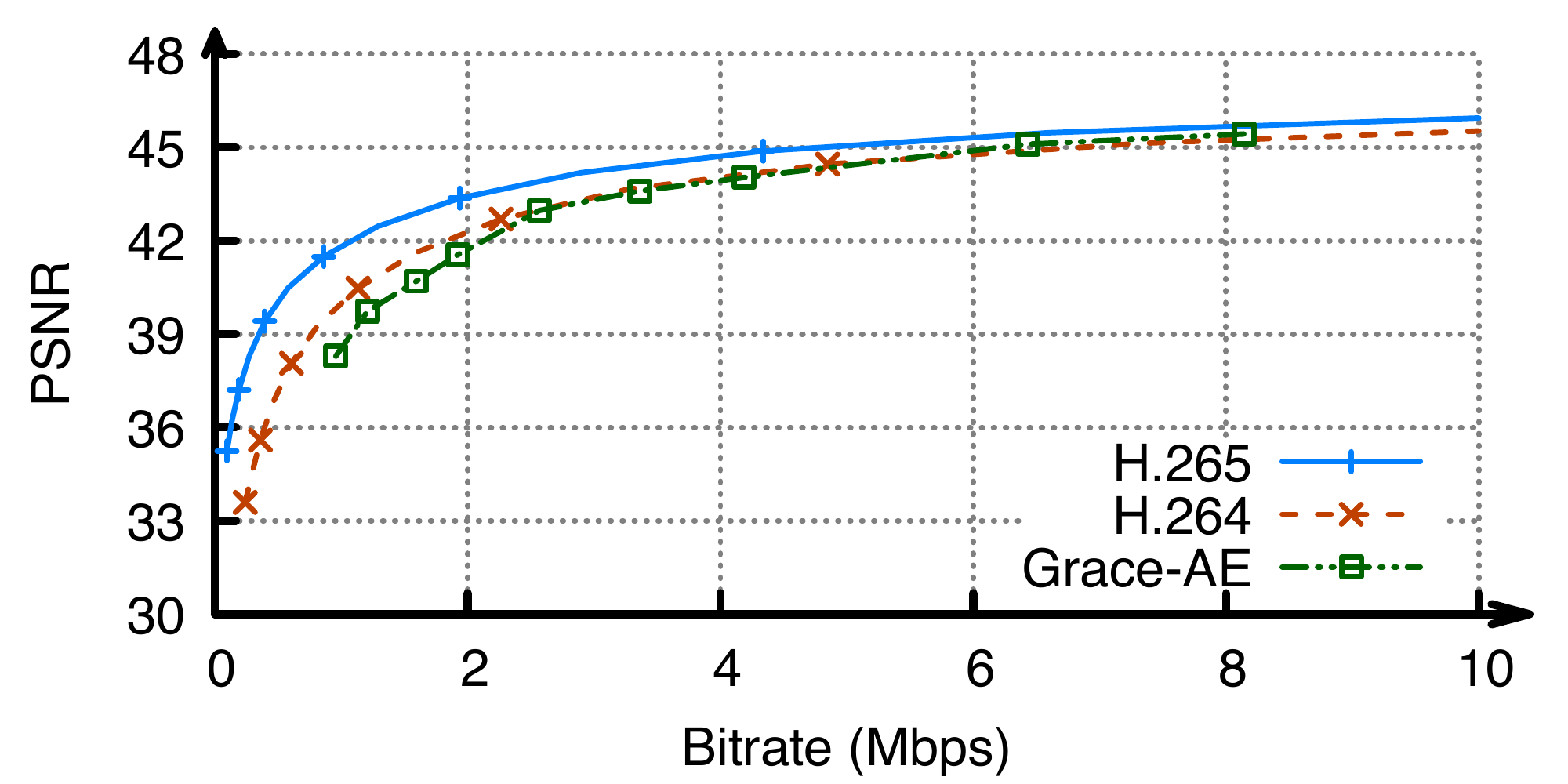}
         \label{fig:}
         \vspace{.05cm}
    \tightcaption{Quality-size tradeoff of \newae. \vspace{.05cm}}
    \label{fig:psnr-vs-bpp}
\end{figure}

For any loss-resilient schemes, the ability to tolerate packet losses often comes with a drop of quality in absence of packet losses, and \newae is no exception. 
Figure~\ref{fig:psnr-vs-bpp} compares the \newae's quality under no packet losses with that of H.264 and H.265 across the test videos (all of which are randomly sampled from three datasets that are different from the training set, \S\ref{subsec:eval:setup}).
% evaluates the price \newae pays to get its loss resilience. It shows the average PSNR of \newae, H.264 and H.265 on different bitrate across all test videos.
Fortunately, at the same encoded bitrate, \newae is similar to H.264 and is only slightly worse than H.265. 
When bitrate varies from 2Mbps to 8Mbps, the average PSNR is 44.29, 44.31, and 44.84 for \newae, H.264 and H.265 respectively.

Remember that H.26x uses the zerolatency option and fast preset (which is typical to real-time videos, including WebRTC\footnote{Google's WebRTC~\cite{WEBRTCSRC} uses OpenH264~\cite{OPENH264}. They set the runtime parameter "usageType" to CAMERA\_VIDEO\_REAL\_TIME  (similar to zerolatency) and "complexityMode" to LOW\_COMPLEXITY (similar to fast)}. %\jc{Yihua, please add a footnote about the following: one is usageType = CAMERA\_VIDEO\_REAL\_TIME  (similar to zerolatency) and another is complexityMode = LOW\_COMPLEXITY (similar to fast)})
We confirm that without these settings, H.265's coding efficiency is better than \newae in the absence of packet losses.
However, our goal is a data-scalable autoencoder for real-time videos (shown shortly), rather than an autoencoder that outperforms heavily engineered H.26x codecs. 
Moreover, the coding efficiency of autoencoder could still be improved (\eg our autoencoder estimates motion vectors based only on one reference frame).

\begin{figure}[t!]
    \centering
     \begin{subfigure}[b]{0.49\linewidth}
         \centering
         \includegraphics[width=2.0\linewidth]{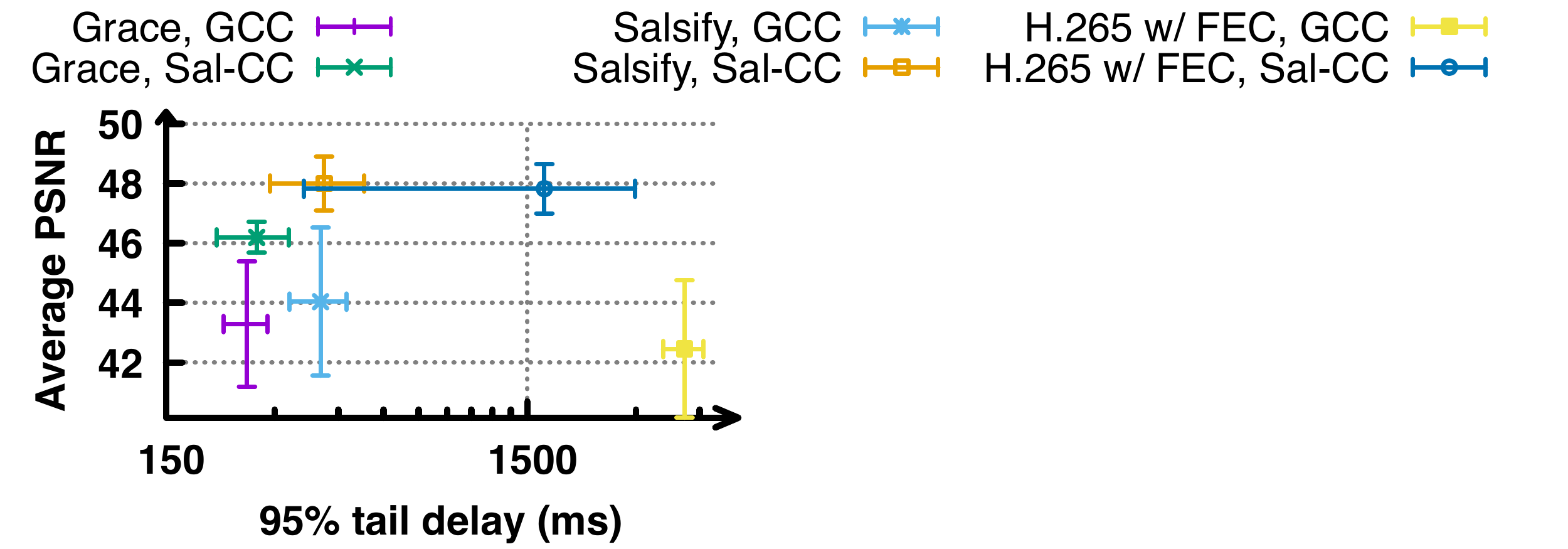}
         \caption{PSNR}
         \label{fig:}
     \end{subfigure}
     %\hfill
     \begin{subfigure}[b]{0.49\linewidth}
         \centering
         \includegraphics[width=1.0\linewidth]{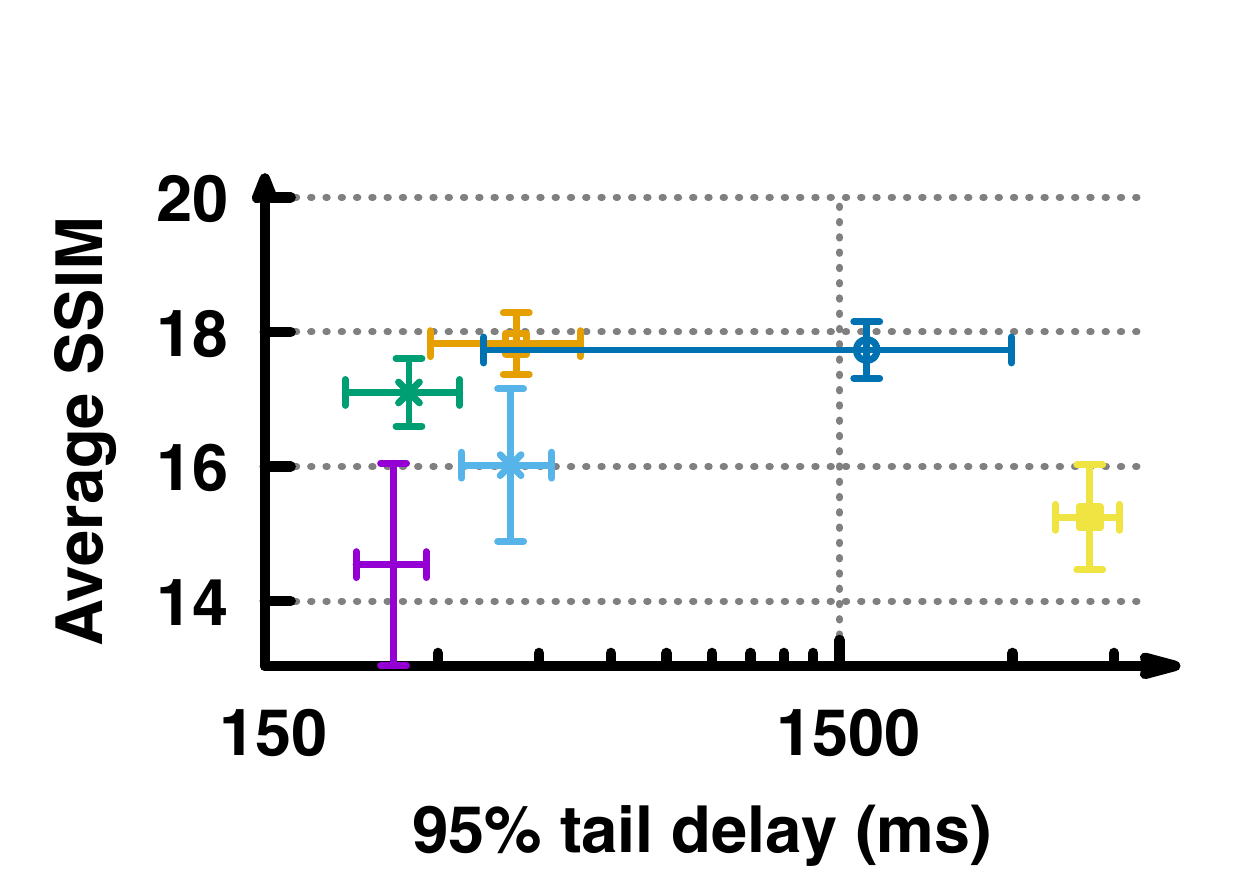}
         \caption{SSIM}
         \label{fig:}
     \end{subfigure}
     \vspace{.05cm}
    \tightcaption{Video quality vs. tail delay (by quality metrics) \vspace{.1cm}}
    \label{fig:e2e-quality}
\end{figure}

\begin{figure}[t!]
    \centering
    \begin{subfigure}[b]{0.49\linewidth}
         \centering
         \includegraphics[width=2.0\linewidth]{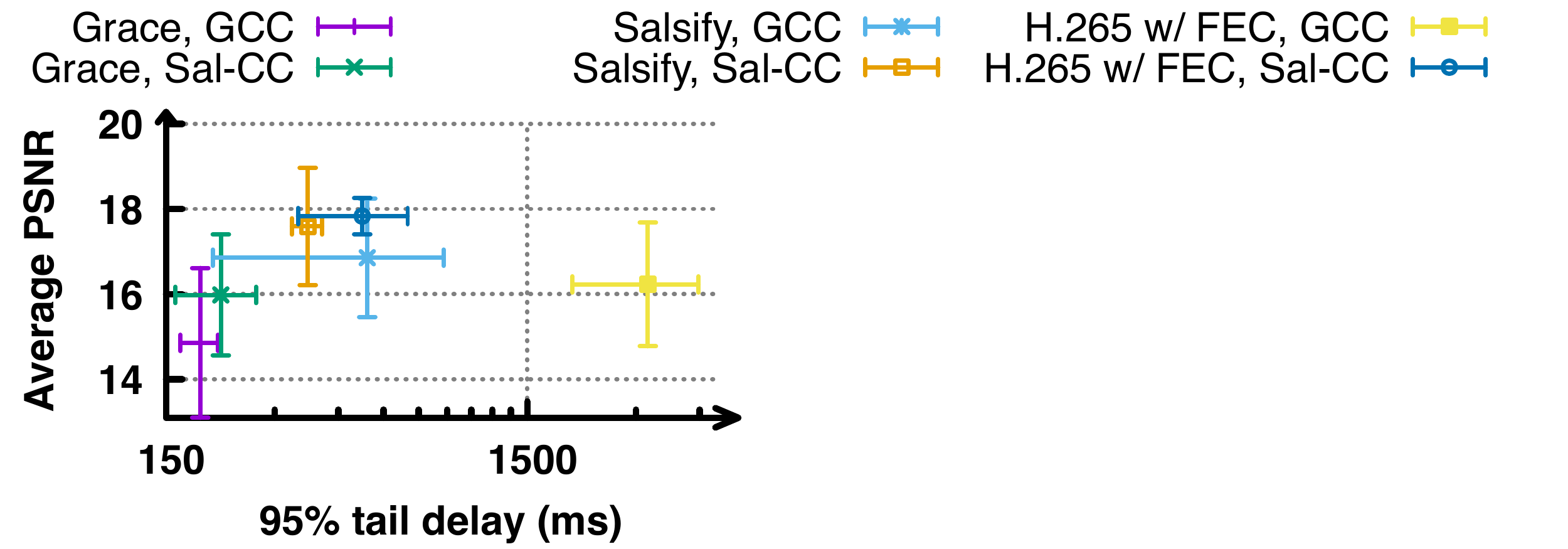}

         \caption{Propagation delay = 100ms)}
         \label{fig:}
     \end{subfigure}
    %  \hfill
    %  \begin{subfigure}[b]{0.32\textwidth}
    %      \centering
    %      \includegraphics[width=\textwidth]{figs/e2e-medrtt.pdf}
    %      \caption{Medium RTT (200ms)}
    %      \label{fig:}
    %  \end{subfigure}
    %  \hfill
     \begin{subfigure}[b]{0.49\linewidth}
         \centering
         \includegraphics[width=1.0\linewidth]{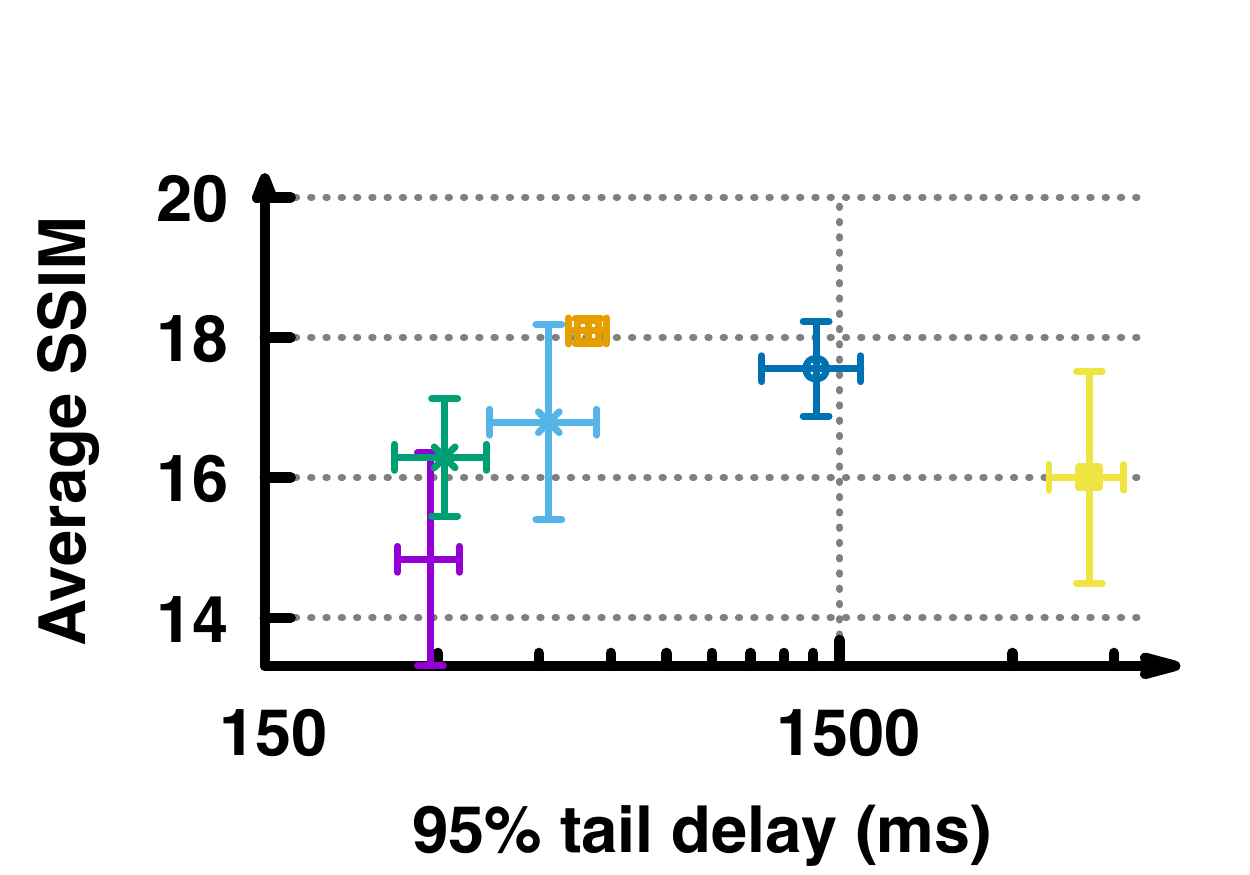}
         \caption{Propagation delay = 300ms)}
         \label{fig:}
     \end{subfigure}
     \vspace{.05cm}
    \tightcaption{Video quality vs. tail delay under different RTT\vspace{.1cm}}
    \label{fig:e2e-rtt}
\end{figure}

\begin{figure}[t!]
    \centering
    \begin{subfigure}[b]{0.49\linewidth}
         \centering
         \includegraphics[width=2.0\linewidth]{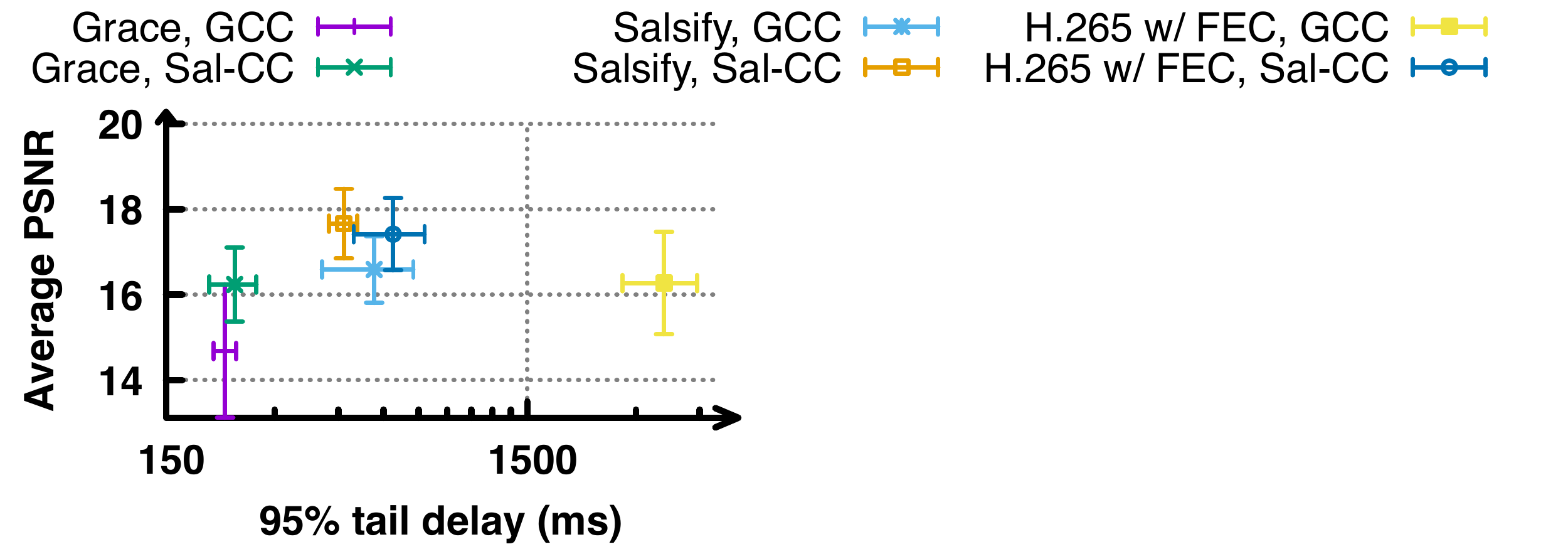}
         \caption{Droptail queue = 15 packets}
         \label{fig:}
     \end{subfigure}
    %  \hfill
    %  \begin{subfigure}[b]{0.32\textwidth}
    %      \centering
    %      \includegraphics[width=\textwidth]{figs/e2e-medqueue.pdf}
    %      \caption{Medium queue (25 packets)}
    %      \label{fig:}
    %  \end{subfigure}
    %  \hfill
     \begin{subfigure}[b]{0.49\linewidth}
         \centering
         \includegraphics[width=1.0\linewidth]{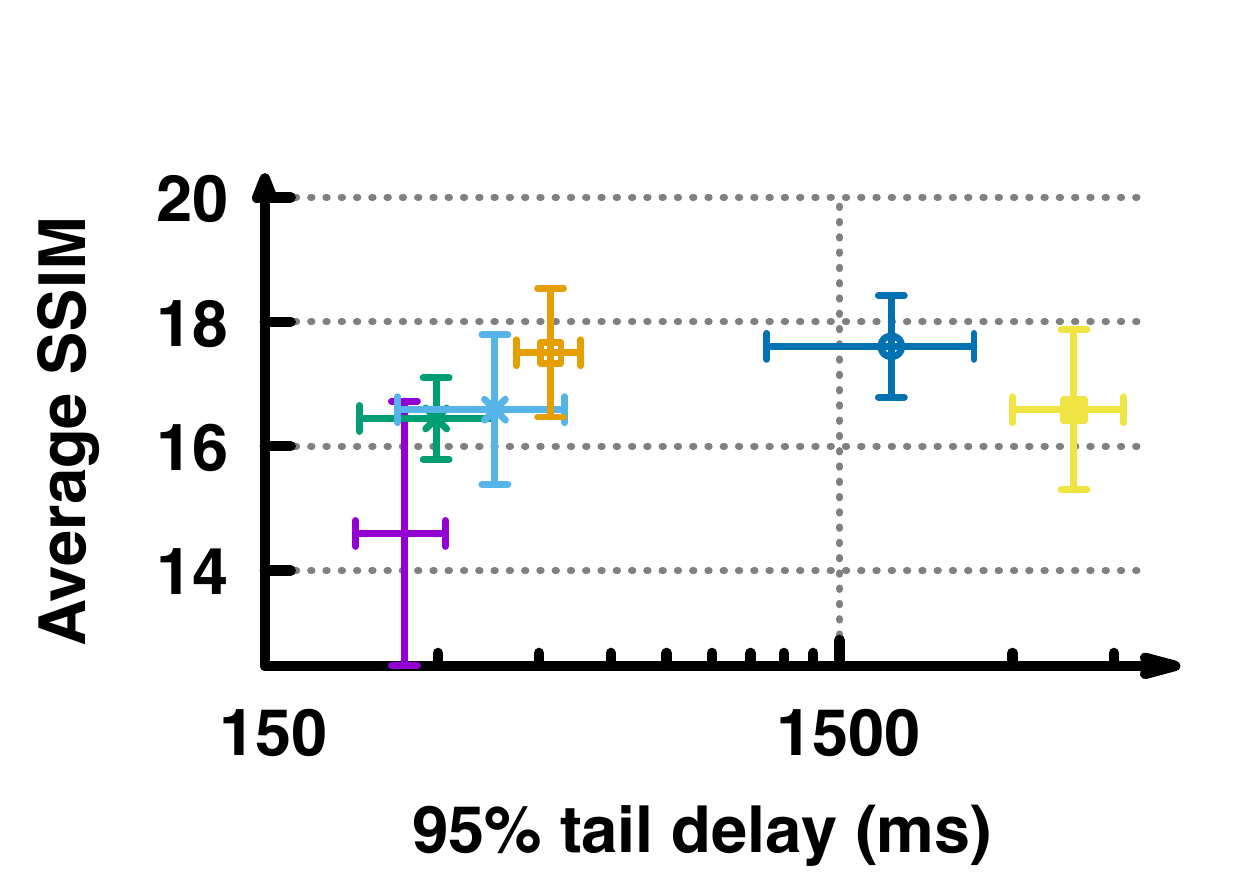}
         \caption{Droptail queue = 35 packets}
         \label{fig:}
     \end{subfigure}\vspace{.05cm}
    \tightcaption{Video quality vs. tail delay under different queues \vspace{.1cm}}
    \label{fig:e2e-queue}
\end{figure}

\begin{figure}[t!]
    \centering
    \includegraphics[width=0.75\linewidth]{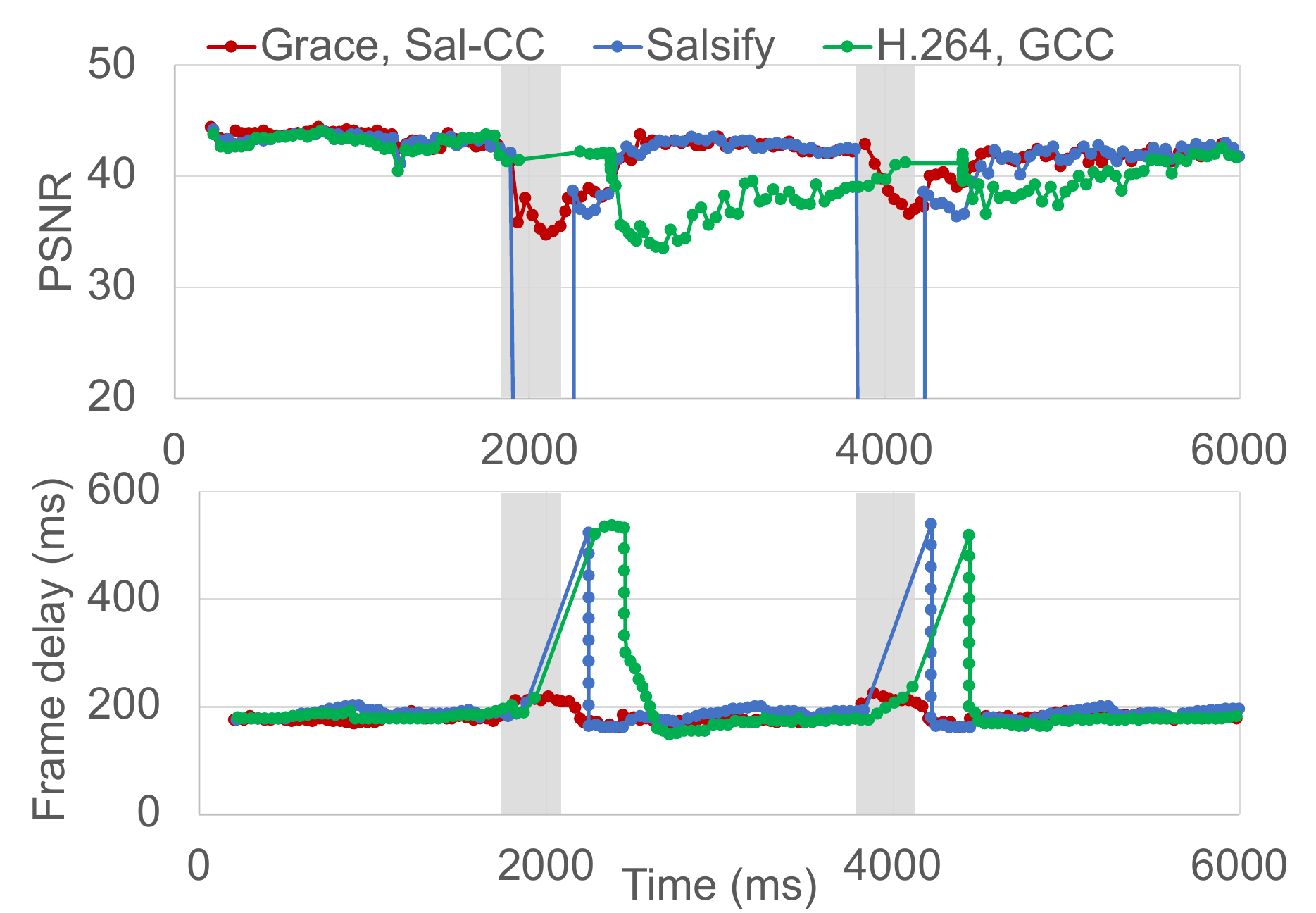}
    \tightcaption{\name achieves lower delay and decent quality during sudden bandwidth drops.}
    \label{fig:timeseries-ours}
\end{figure}

\tightsubsection{Video quality vs. tail frame delay}
\label{subsec:eval:e2e}

Next, we run \name and the baseline codecs on real bandwidth traces from Verizon and T-Mobile LTE (\S\ref{subsec:eval:setup}). 
We set RTT at default values (200ms) and droptail queue length of 25 packets.
We implemented two different congestion control algorithms: Google congestion control~\cite{carlucci2016analysis} (\textit{GCC}) and the congestion control from the previous work (\textit{Sal-CC})~\cite{salsify}. 
We also implemented FEC-protected H.265 and Salsify codec as the baseline codecs (\S\ref{subsec:eval:setup}).

Figure~\ref{fig:e2e-quality} compares \name and the baselines, in average quality (both PSNR and SSIM) and the 95-th percentile of frame-level delay across frames in a session. 
We see that \name significantly decreases the tail frame delay: \name is 251ms and 267ms when using GCC and Sal-CC respectively, compared to 401ms and 412ms for Salsify codec, and 4066ms and 1652ms for H.265 codec. When there is a packet loss, Salsify codec needs to skip frames, causing a delay increase. For H.265 codec, it needs to wait for those lost packets being retrasmitted, and such retransmission traffic will compete with the normal video traffic, causing a higher delay on H.265.

When under the same congestion control (Sal-CC or GCC), the average PSNR of \name, Salsify codec and H.265 are 46.2, 47.8, 47.9, respectively. 
Though \name's PSNR is on par or slightly lower, we believe the reduction in tail frame delay is worth the minor drop in quality.
We make a similar observation when measuring quality in SSIM (dB): \name's quality is similar or slightly lower than the quality of baseline using the same CC.

% \yihua{check if we already mentioned congestion control before}
%We play a \fillme-second long video on \fillme cellular network traces~\cite{??} and set the propagation delay at \fillme ms and queue length at \fillme. 
%\jc{YIHUA, please fill in some detailed reading}

%\jc{
%\begin{itemize}
%    \item by trace set
%    \item by rtt
%    \item by queue length
%    \item a concrete example
%\end{itemize}
%}

We then test \name's performance on the same traces but under different queue lengths and different propagation delays.
Figure~\ref{fig:e2e-rtt} and Figure~\ref{fig:e2e-queue} show that \name consistently reduces tail delay under different network delays or under different droptail queue lengths. 
In particular, \name has more delay reduction when the queue is shorter. 
This is because during congestion, shorter queues tend to drop previous frames packets more aggressively, and might still delivery {\em some} packets (though not all) of the latest frame to the receiver. 
In such cases, \name lets the client decode the frame early based on any subset of packets it has received. 

% we use synthetic bandwidth traces, where the bandwidth is 5Mbps but scattered with episodes of low bandwidth 1Mbps for 500ms (see Figure~\ref{??} for an example) and these low bandwidth episodes come at a Poisson distribution. 
% Such patterns are typically used in prior work (\eg~\cite{??,??) } to emulate impact of bursty competing flows.

% As shown in Figure~\ref{??}, \name has higher tail delay reduction under higher propagation delays or shorter queues. \jc{YIHUA, please fill in some detailed reading}
% These results highlight that \name reduces delay when packet retransmission or redundancy-based schemes do not suffice. 
% Likewise, our gain will be negligible if the RTT is very low to allow instant packet retransmission (in which case, we speculate SVC will be more fitting), or if the packet losses are largely constant (in which case, FEC will be an ideal choice).

Finally, we show how \name and the baselines behavior in a concrete bandwidth trace sample (Figure~\ref{fig:timeseries-ours}).
The bandwidth drops from 5Mbps to 1Mbps at 1.9sec and last for 200ms, before bouncing back to 5Mbps (another bandwidth drop occurs at 3.9sec).
We can see that during each bandwidth drop, \name's delay does not increase as sharp as the baselines (notably Salsify is the second best since it could skip frames).
\name and Salsify use the same CC, so their quality roughly matches on frames not skipped by Salsify.
That said, \name's quality degrades marginally with no frame skipping during congestion, and it bounces back once packet losses disappear.

% \jc{YIHUA, please fill in some detailed reading}

\begin{figure}[t!]
    \centering
    \begin{subfigure}[b]{0.49\linewidth}
         \centering
         \includegraphics[width=\linewidth]{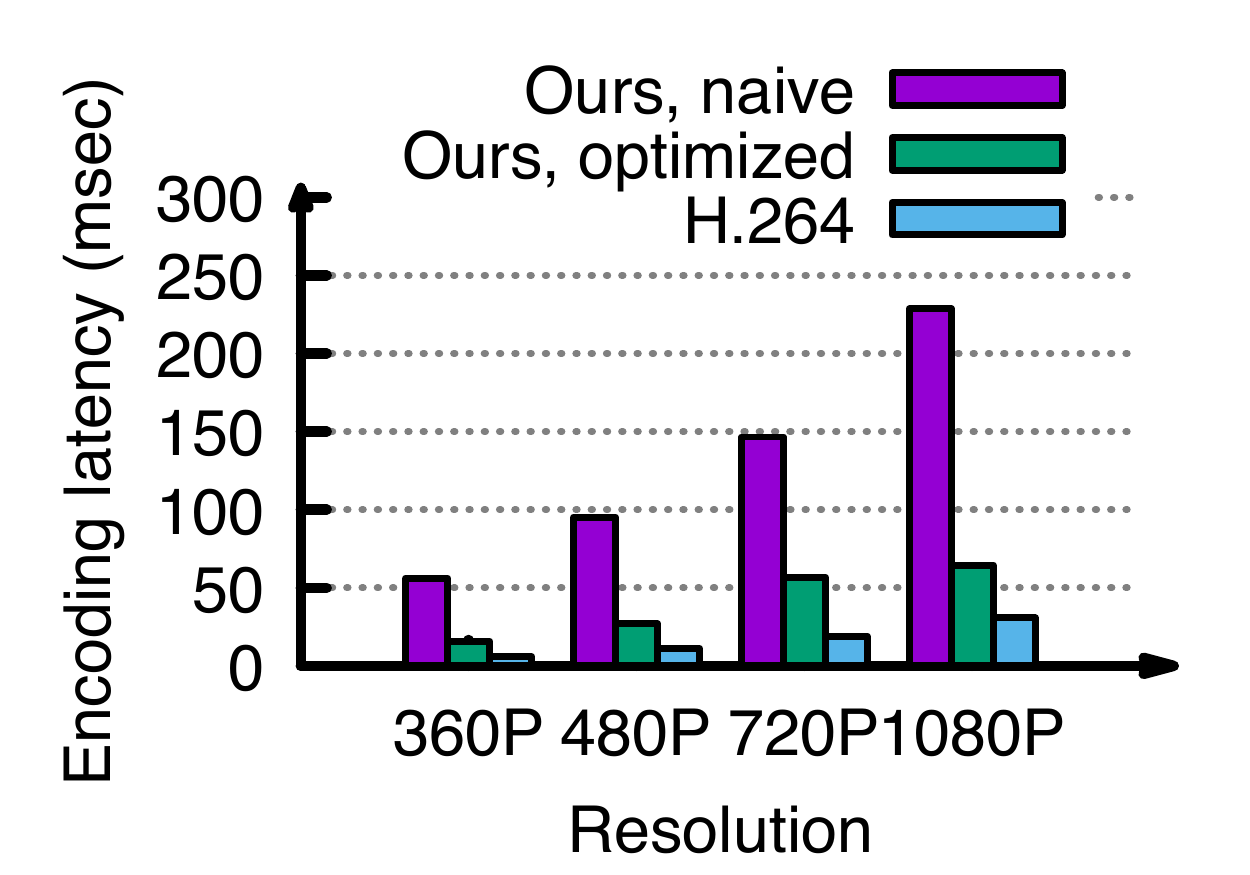}
         \caption{Encoding latency}
         \label{fig:system-encoding}
     \end{subfigure}
     \hfill
     \begin{subfigure}[b]{0.49\linewidth}
         \centering
         \includegraphics[width=\linewidth]{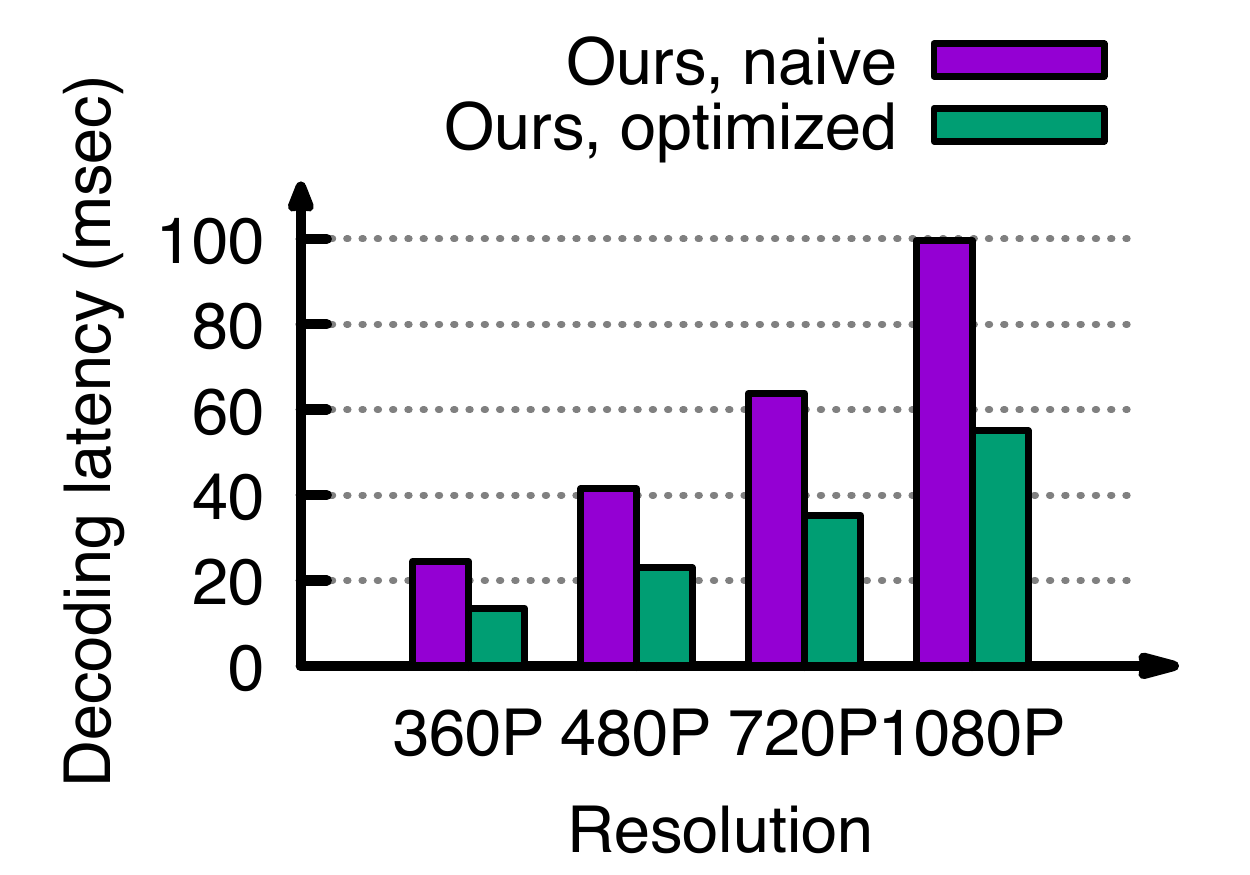}
         \caption{Decoding latency}
         \label{fig:system-decoding}
     \end{subfigure}
     \hfill
     \begin{subfigure}[b]{0.49\linewidth}
         \centering
         \includegraphics[width=\linewidth]{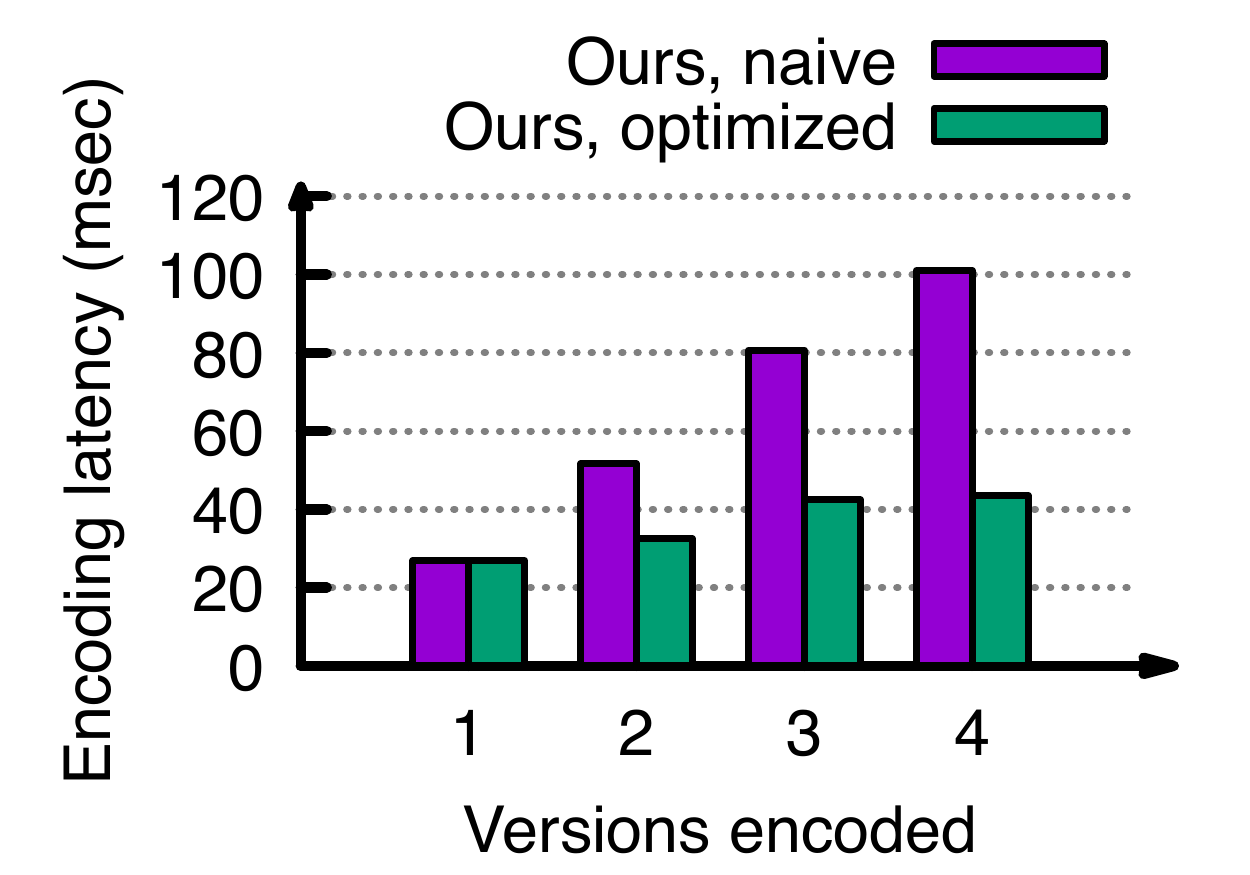}
         \caption{Latency of encoding multiple versions}
         \label{fig:system-encoding-versions}
     \end{subfigure}
     \hfill
     \begin{subfigure}[b]{0.49\linewidth}
         \centering
         \includegraphics[width=\linewidth]{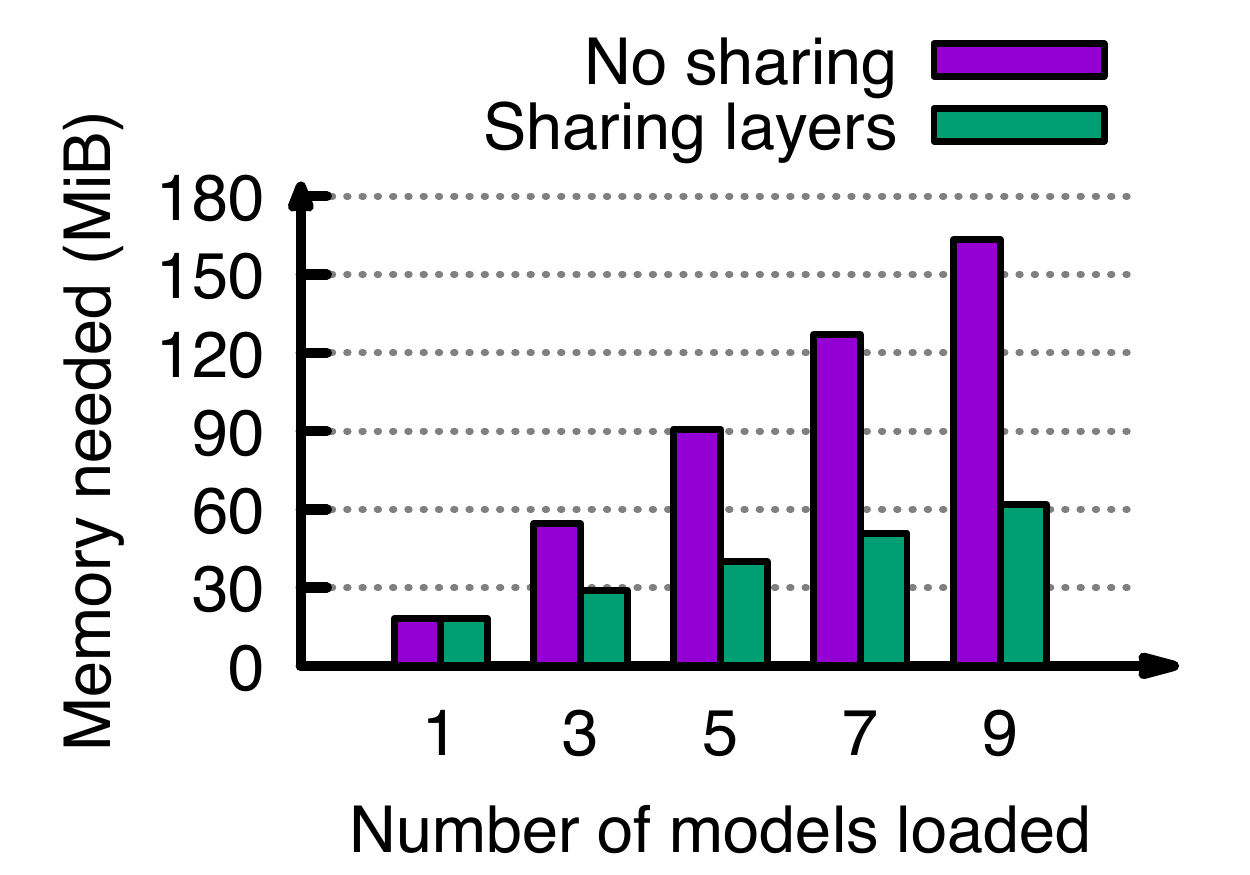}
         \caption{Memory usage of loading multiple models}
         \label{fig:system-memeory}
     \end{subfigure}
     \hfill
     \tightcaption{Memory and compute overhead of \newae}
     \label{fig:}
\end{figure}

\tightsubsection{Memory and computing overhead}
\label{subsec:eval:system}

% \zy{There is a typo ("decodeing latency") on the y-axis label on Figure 16 (b). (figs/decode.pdf)}

% Remember that \newae needs a P-frame encoder and a I-frame encoder to encode each frame as a P-frame and an I-patch (a small spatial region without reference).
Our implementation of \newae optimizes both memory footprint and encoding/decoding delay (described in \S\ref{subsec:optimize}), and we test its effectiveness on a Nvidia GeForce RTX 3080 GPU. 
Figure~\ref{fig:system-encoding} and Figure~\ref{fig:system-decoding} show that compared to the original (unoptimized) implementation of pre-trained autoencoder (which takes 100ms to encode a 480p frame), our implementation can encode a 480p frame in less than 25ms (a $4\times$ speedup).\footnote{The only autoencoder that achieves comparable encoding/decoding speed is~\cite{elf}, and it achieves such speed on a 5x more expensive Titan V GPU. We hope to compare it with \newae, but it is not open-source.}
This translates to almost 40fps. 
Even for 720p HD videos, our optimized codec can achieve 18fps (55ms encoding per frame). 
% \fillme~fps/\fillme~fps, \newae encodes/decodes 480p frames at \fillme~fps/\fillme~fps, a \fillme$\times$ speedup.
This acceleration comes from downsampling frames in motion-vector estimation. 
Moreover, when adapting to a target bitrate, \name encodes the next frame multiple times {\em without} needing to run the entire encoder multiple times, because it re-uses the motion-compensated prediction, so only the residuals need to be encoded at multiple bitrates.
Figure~\ref{fig:system-encoding-versions} shows that the delay of encoding a frame at three different bitrates only increases marginally compared to encoding one frame. 

In terms of memory usage, our optimized implementation saves the memory footprint of loading 9 models by over 60\% since most NN layers are shared. 
% Figure~\ref{??} shows that compared to loading the multiple autoencoders (I-frame encoder/decoder and a P-frame encoder/decoder of each bitrate) in GPU memory, the memory usage \name's autoencoders (not including the memory usage of intermediate features) is \fillme\% smaller, because they share most of the layers (only the last \fillme layers of the encoder and the first \fillme layers of the decoders are different). 
Note that \newae does not change the NN architecture of the autoencoders---only the weights are retrained and the way they are used is optimized. We speculate that they could be made more efficient with appropriate modification to its NN architecture.

% Figure~\ref{??} validate that \newae's coding efficiency is comparable to (even slightly better than) without the optimizations (\ie without sharing the NN backbone or downsampling and sharing the motion-vectors.)
% This is because, though quality does drop with these optimizations, the encoded size also drops (due largely to the downsampled motion vectors), which result in a similar bitrate-quality trade-off frontier.

% \jc{The only autoencoder that achieves comparable encoding/decoding speed is Elf~\cite{??}. 
% We hope to compare it with \newae, but it is not open-source. }

% \jc{remember to highlight that the speed is comparable to elf, which runs on the 5x more expensive titan v gpu.}

\begin{comment}

\subsection{Design choices in \name}

\jc{
\begin{itemize}
    \item different training loss patterns
    \item different training \rate distributions
    \item saliency-based redunancy
    \item packet prioritization
    \item computing overhead in delay and in memory
\end{itemize}
}

\end{comment}
%\input{sections/eval_v2}

%!TEX root = ../main.tex
%!TEX spellcheck = en_US

\tightsection{Related work}
\label{sec:related}

% \begin{itemize}
% \item loss resilient schemes
% \item adaptive streaming: cc, fec rate, bitrate, etc.
% \item autoencoders
% \end{itemize}

% Other than the prior work elaborated elsewhere, we want to briefly mention a few other related techniques.

\mypara{Loss resilience}
Loss resilience techniques fall into three categories. 
Channel coding includes FEC techniques, Reed-Solomon code, LDPC and fountain codes~\cite{mackay1997near,mackay2005fountain}.
%An ideal fountain code sends as many bytes as possible and can reconstruct the original file once same number of bytes of the original file have been received. 
While the (ideal) ability to reconstruct video of bitrate $k$ from any received data of bitrate $k$ is appealing, it means that some packets must be delayed/retransmitted if the encoded video bitrate already exceeds the link capacity, so using channel coding alone is not ideal for real-time video applications~\cite{fong2019low}.
Source coding adds redundancy during the video coding process, such as Intra-MB insertion~\cite{cote2000optimal}. 
% which encodes some macroblocks (MBs) in a P-frame without referring to previous frames, making them decodable in absence of the reference frame.
A more promising approach to loss-resilient video coding is joint-source-channel coding, and our \newae belongs to this category.
The closest related work is DeepJSCC~\cite{kurka2020deepjscc} which trains an autoencoder to code images in a representation that is both compact and robust to signal noises~\cite{bourtsoulatze2019deep,gunduz2019machine}.
%It combines data (image) compression with robustness to low SNR in the physical layer
\newae differs from DeepJSCC on two key fronts. 
First, they target physical-layer protection against signal noises.
% is different from (and complementary to) transport-/application-layer protection against packet losses. 
While physical-layer noises can be modeled as differentiable linear transformations (as in~\cite{bourtsoulatze2019deep,gunduz2019machine}), packet drops (\rate) are not differentiable and need to be handled differently (see \S\ref{subsec:sim-losses}). 
Second, unlike compressing individual images, frames in a streaming video cannot be treated separately, otherwise any error on one frame may propagate to future frames. 
%While more research is needed, we believe that we have taken a significant step forward in showing the promise of autoencoders in real-time videos and outlining the concrete technical steps that should be pursued. 

\mypara{Video compression}
Obtaining loss resilience must sacrifice coding efficiency (lower quality in absence of packet losses), and \newae is no exception.
Most commercial video coding techniques are based on classic codecs, such as H.265 and VP9. 
Though video autoencoders (\eg~\cite{dvc,elf}) have greatly improved their coding efficiency, it remains unclear if they outperform heavily engineered state-of-the-art video codecs, which do not limit the search for optimal reference MB and can run multiple passes on a video to optimize coding efficiency. 
Our evaluation focus on a specific setting (H.26x without B-frames and the ``fast'' preset) and shows that \newae's coding efficiency is comparable with H.26x.
In a recent work~\cite{swift}, autoencoders' coding efficiency is also shown to be comparable with H.265. 
That said, we choose to build \name on autoencoders not for its coding efficiency, but for its potential to realize data-scalable delivery.

\mypara{Adaptive video streaming}
There has been intense research on adaptive streaming, including adapting sending rate (\eg~\cite{sprout,concerto}), encoding bitrate (\eg~\cite{salsify,xiao2018two,zhang2021loki}) and FEC redundancy rate (\eg~\cite{nagy2014congestion,fong2019low,holmer2013handling}).
These congestion control and bitrate adaptation algorithms are essentially reactive to intermittent congestion and packet losses. 
\name is complementary to them in that data-scalable delivery eschews the need for accurate loss prediction and instead gracefully degrades their quality with more packet losses.
% as the tail delay is susceptible to occasional packet losses/delaying or bandwidth drops. 
% While they achieve good performance with accurate predictions, it not easy to always be accurate, and thus the tail quality and delay is susceptible to occasional packet losses/delaying or bandwidth drops. 
%In contrast, data-scalable coding is complementary to rate adaptation, but they eschew the need for accurate loss prediction and instead gracefully degrade their quality with more packet losses.

\tightsection{Limitations}

\mypara{At least one packet has to arrive}
\name allows the receiver to choose when to decode {\em after} the first packet arrives, but it cannot decode a frame with no received packet. 
This means \name will not help during network disconnection or severe congestion that blocks any packet arrivals.
One potential way to avoid  such extreme situation is to leverage multiple concurrent connections and run \name over multipath TCP (with custom packet scheduling to discard or deprioritize untimely packets.)

\mypara{Bounded by autoencoder's performance}
Though \S\ref{sec:eval} shows that \newae's compression efficiency is comparable to H.264 under no packet loss, its quality-bitrate tradeoff is bounded by that of the pre-trained autoencoders. 
This is expected, since we choose not to modify the autoencoder architecture, and instead only train them to be data-scalable. 
Similarly, \newae does not improve the generalization of pre-trained autoencoders on unseen videos, because we train them using the pre-trained models' training data (Vimeo-90K~\cite{vimeo-dataset} and COCO\cite{coco-dataset}).
That said, autoencoders are being constantly improved to obtain higher compression efficiency (\eg via using more reference frames like in traditional codecs) and work better on new videos (\eg via model adaptation).

%\mypara{No multi-party support (yet)}

%\mypara{First I-frame must be lossless}

\mypara{Frame skipping}
\name rarely skips a frame, unless no packet is received at all.  
This is a unique benefit of data-scalable delivery. 
However, we acknowledge that this is not necessarily the optimal choice.
We believe more user studies are needed to understand if people prefer frames decoded at a slightly lower quality (due to packet losses) or skip a few frames in hope of improving the quality of future frames.

\mypara{Compute and power overhead}
\name belongs to the line of work (\eg~\cite{nas,swift}) that embraces the trend of more hardware NN accelerators. 
That said, \newae as-is runs much more slowly than the heavily engineered codecs, so it is not applicable to low-power devices.

\tightsection{Conclusion}
This paper presents \name, a new real-time video system based on autoencoders trained to deliver video frames in a data-scalable manner (decodable with any non-empty subset of packets and graceful improvement with more packets).
Though in absence of packet losses, \name has lower coding efficiency than traditional codecs, we show that \name allows clients to better balance video quality and frame delay.

%-------------------------------------------------------------------------------
\bibliographystyle{plain}
\bibliography{reference}

\appendix

\section{Details of autoencoder architecture and training}

%\jc{Anton, please add the model architecture figure here and some description of the layers, input output, and how we train the model (training rate, batch size, training data, etc)}

\subsection{I-frame compression model}

\begin{figure}[!htb]
    \centering
         \centering
        \includegraphics[width=0.99\linewidth]{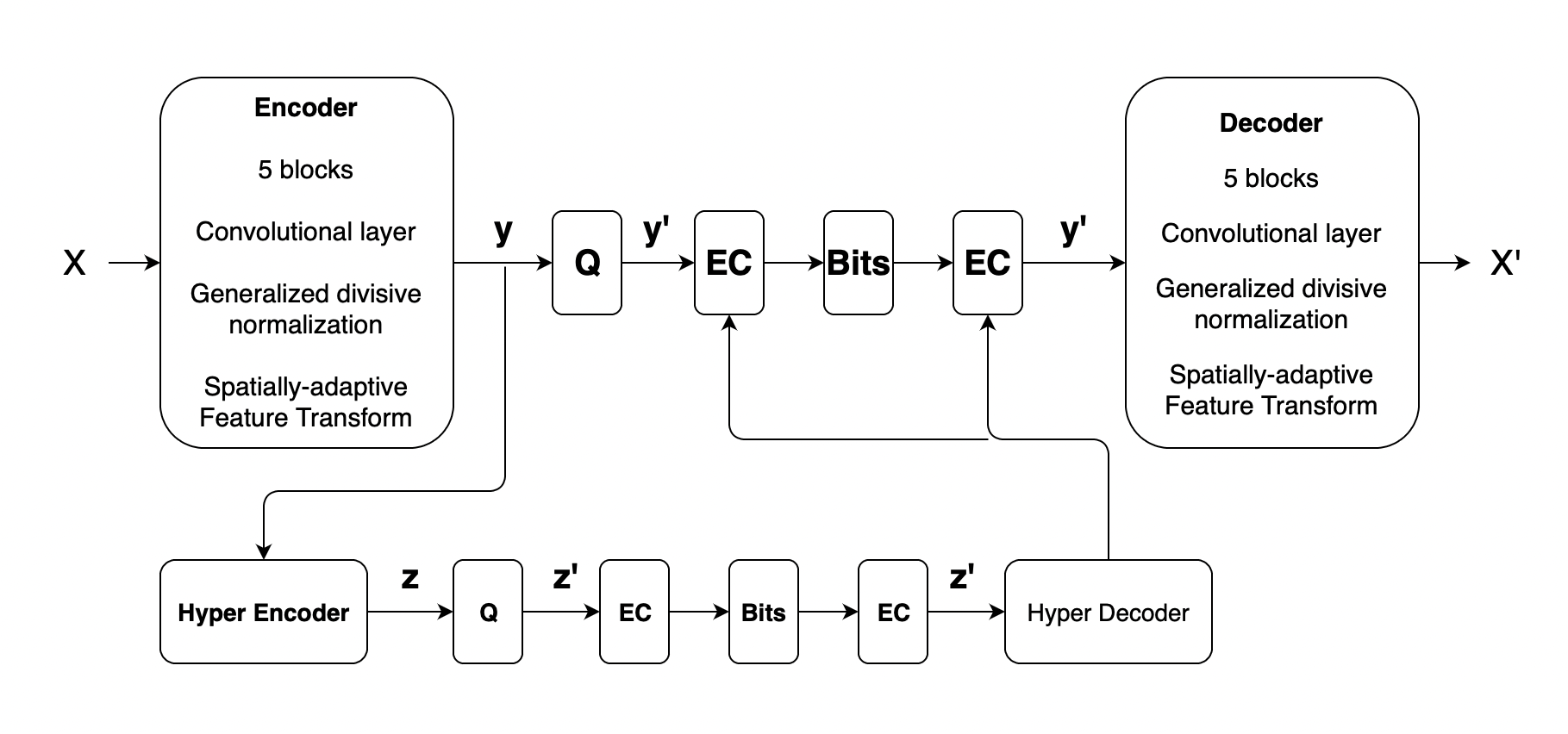}
         \label{fig:}
    \caption{High-level architecture of I-frame compression autoencoder.}
    \label{fig:}
\end{figure}

Autoencoder architecture for compressing I-frames consists of an encoder, a decoder, a hyper-encoder, and a hyper-decoder networks.

Encoder network is used to generate a latent representation of the input image. It consists of five blocks and a fully connected layer in the end. Each block consists of a convolutional layer, generalized divisive normalization layer, and a Spatial Feature Transform Layer. The last fully connected layer serves as a converter resized for the latent representation size. It can be easily fine-tuned by freezing the rest of the layers and re-training only the fully connected layer for a short time to fit the desired size for the latent representation of the image.

Hyper-encoder and hyper-decoder networks are used to generate side-information from the latent representation of the image. The output of the hyper-encoder is used as an input to hyper decoder to generate the parameters (mu, sigma) for the Gaussian entropy model, which approximates the distribution of the latent representation of the image. The hyper-encoder consists of three blocks, where each block consists of a convolutional layer, Spatial Feature Transform layer, and a LReLU. The hyper-decoder consists of three blocks, each of which consists of a convolutional layer and a LReLU.

Decoder uses the reconstructed latent image representation and the reconstructed side-information generated by the hyper-encoder to reconstruct the image. Decoder consists of five blocks, each of which consists of convolutional layer, a generalized divisive normalization layer, and a Spatial Feature Transform layer.

We apply quantization to both latent image representation and side-information generated by hyper-encoder before they make it into the decoder and hyper-decoder respectively.

Autoencoder model to compress I-frames was trained on the COCO Dataset consisting of COCO 2014 and COCO 2017 which sum up to 201,000 images in total. We used the batch size of 8, with learning rate of 0.0001. It took 1,980,000 iterations to train the base model with 0\% packet loss, and 100,000 iterations to fine-tune the model with packet loss.

\subsection{P-frame compression model}

DVC model to compress P-frames mimics the mpeg encoder and decoder structure, with residual, motion vector encoder and decoder, and motion compensation substituted with convolutional networks.

% residual network

Motion estimation is realized in three steps. First, we use optical flow estimation using a library to estimate motion vectors from optical between two frames. Then, optical flow is encoded using a motion vector encoder. The motion vector encoder consists of four convolutional layers, separated by three generalized divisive normalization layers. After the encoder, we apply quantization and loss to the motion vector data, which then is fed into the decoder. The motion vector decoder consists of four deconvolution layers, with three inverse generalized divisive normalization layers.

We take the output of the motion vector decoder and output of residual decoder and feed it into the motion compensation network. Motion compensation network first takes in the reconstructed motion vector and a reference frame, where reference frame is warped using the reconstructed motion vector. Then, the resulting warped frame together with the reference frame and reconstructed motion vector are fed into a convolutional network, which consists of six convolutional layers with average pooling steps in between them. The output of this convolutional network is the reconstructed frame.

DVC model to compress P-frames was trained on the 90k Vimeo Dataset, with batch size of 4, learning rate of 0.0001 and learning rate decay of 0.1. To control bitrate, we scaled the PSNR component in the loss term by a constant factor $\alpha$, making it larger in proportion to BPP. Furthermore, for the purposes of saving memory that models occupy and training speed, we chose model with $\alpha=1024$ as the "backbone" and fine-tuned only 10\%-30\% of layers to adjust to other bitrates. This means that all models share 90\%-70\% of the layers, with only the last one to two convolutional layers of the residual encoder, last one to two convolutional layers of the motion vector encoder, first one to two convolutional layers of residual decoder and first one to two convolutional layers of the motion vector decoder being trained. This allowed us to save over 2-3$\times$ in model storage memory, as well as for each bitrate model to converge within 10,000 iterations with the application of packet loss.

\subsection{Distribution of video content complexity}\label{app:siti}

\begin{figure}[!htb]
    \centering
         \centering
        \includegraphics[width=0.75\linewidth]{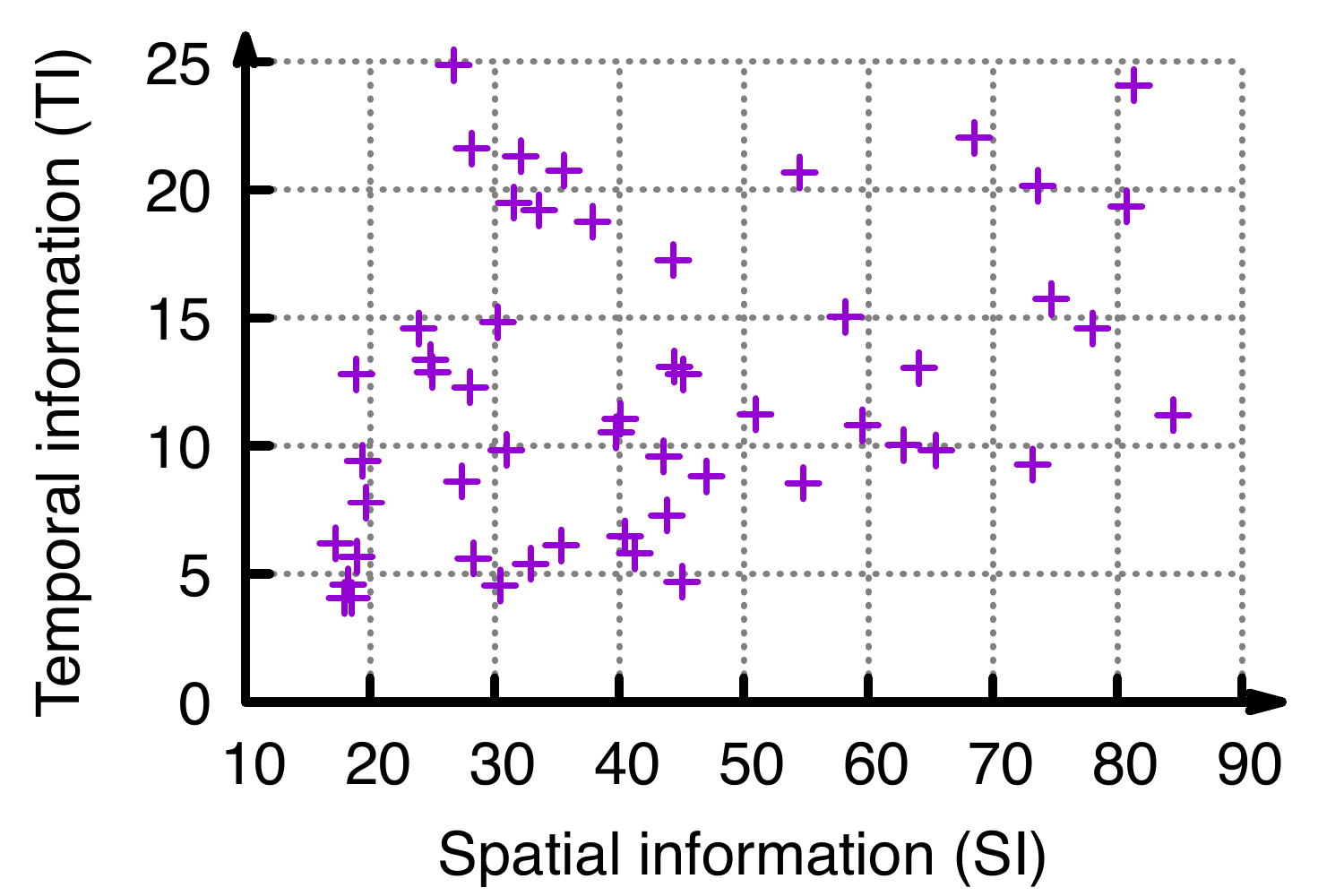}
    \caption{Spatial information (SI) and temporal information (TI) of test videos}
    \label{fig:SITI}
\end{figure}

To validate the test videos that we use cover different content complexities and movements, we calculate the spatiotemporal complexity of the video. We use Spatial Information (SI) and Temporal Information (TI)~\cite{itu1999subjective}, which are frequently-used metrics to measure the spatiotemporal complexity and a larger SI/TI means that the video has a higher spatial/temporal complexity. The metrics are calculated by the tool~\cite{SITITOOL} provided by Video Quality Experts Group (VQEG) and the result is shown in Figure~\ref{fig:SITI}. 

The result validates that {\em (i)} the spatiotemporal complexity of the videos we used covers a wide range: SI is ranging from 15 to 85 and TI is ranging from 3 to 25. %, which corresponds to a recent report.
{\em (ii)} Our test videos covers all the following types: high spatial complexity and high temporal complexity, high spatial complexity but low temporal complexity, low spatial complexity but high temporal complexity, and low spatial complexity and low temporal complexity.

\subsection{Visualization of \newae's Decoded Frames}

\begin{figure*}[!htb]
    \centering
     \begin{subfigure}[b]{0.49\linewidth}
         \centering
         \includegraphics[width=\linewidth]{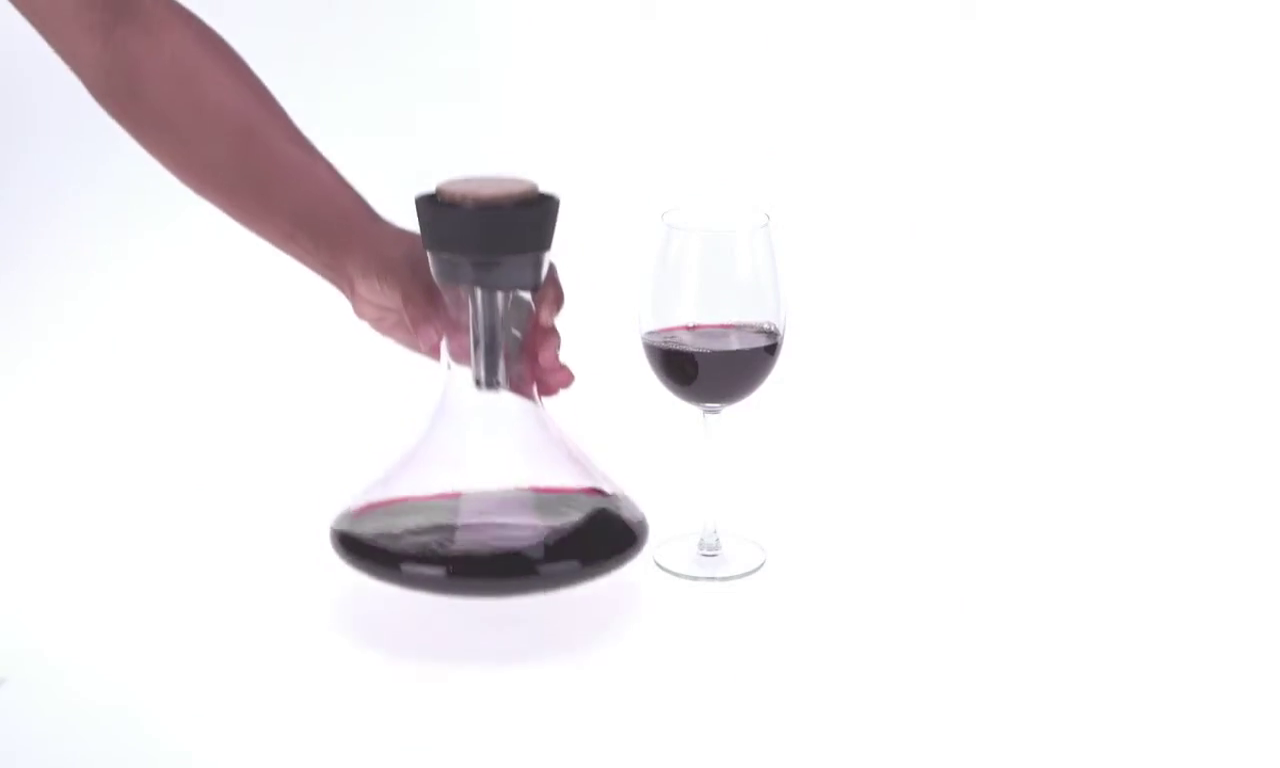}
         \caption{Original frame (SSIM)\\\xspace}
         \label{fig:example-original}
         \vspace{0.3cm}
     \end{subfigure}
     \hfill
     \begin{subfigure}[b]{0.49\linewidth}
         \centering
         \includegraphics[width=\linewidth]{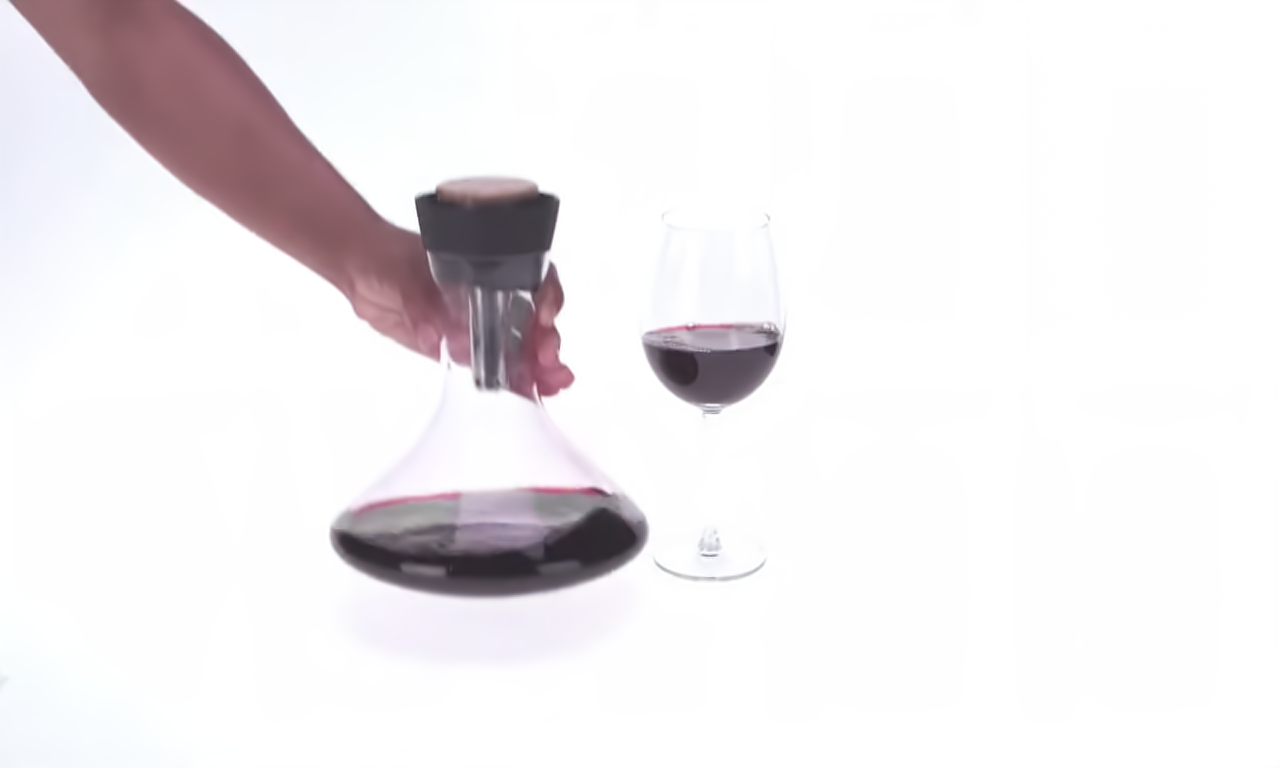}
         \caption{Pretrained autoencoder, no loss (0.9952)}
         \label{fig:example-pretrained-noloss}
     \end{subfigure}
     \hfill
     \begin{subfigure}[b]{0.49\linewidth}
         \centering
         \includegraphics[width=\linewidth]{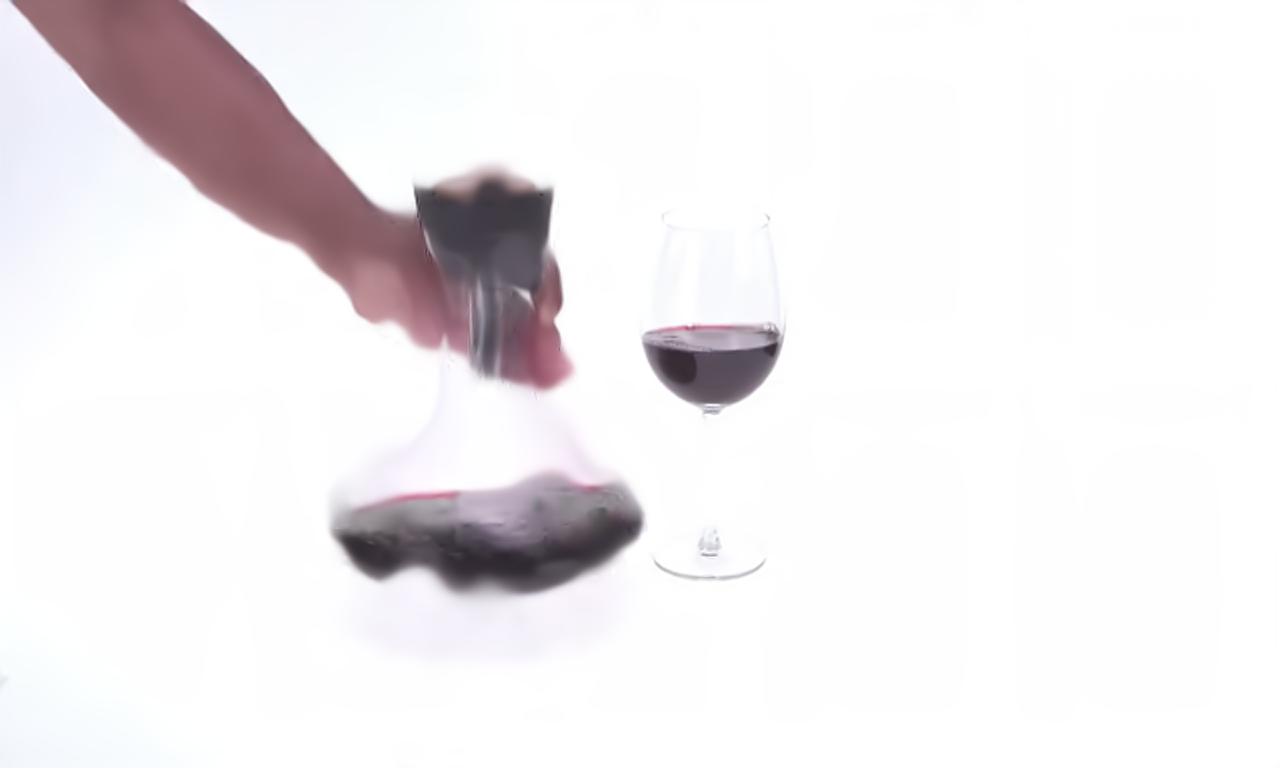}
         \caption{Pretrained autoencoder, with loss (0.9630)}
         \label{fig:example-pretrained-withloss}
     \end{subfigure}
     \hfill
     \begin{subfigure}[b]{0.49\linewidth}
         \centering
         \includegraphics[width=\linewidth]{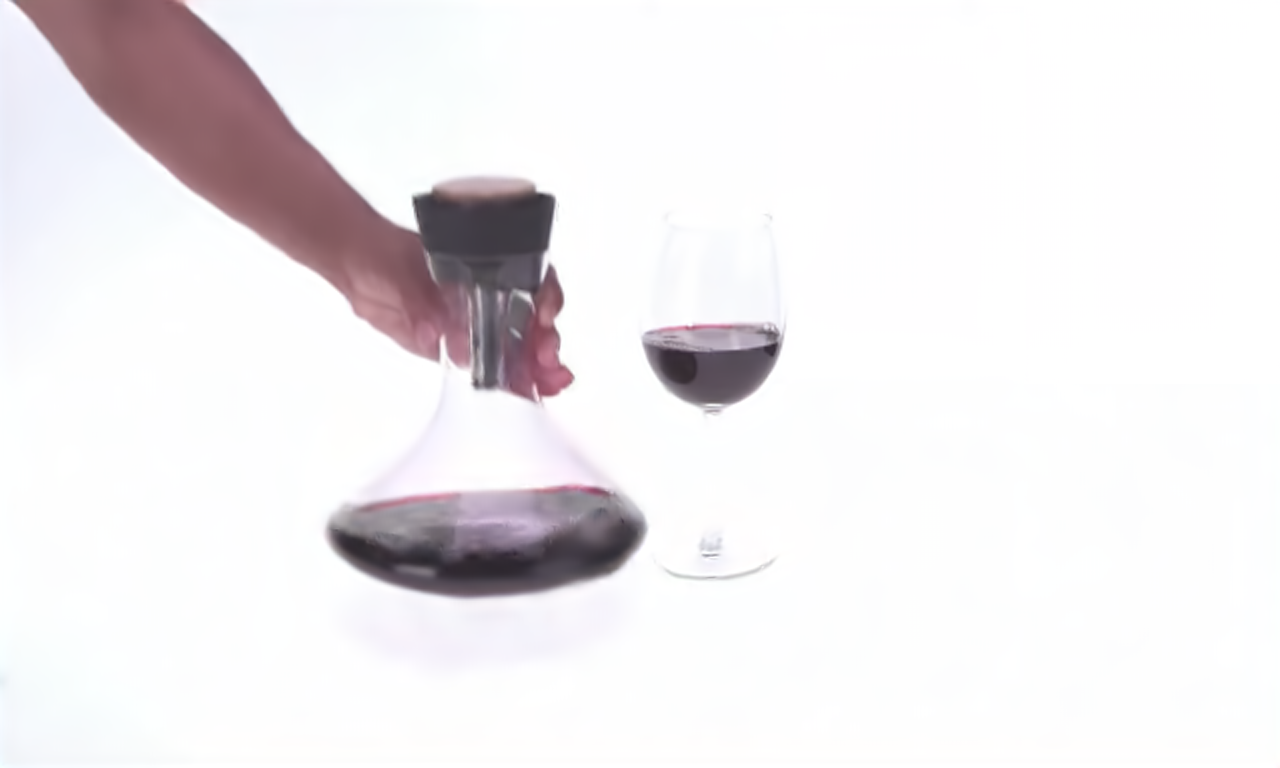}
         \caption{\newae, with loss (0.9830)}
         \label{fig:example-ours-withloss}
     \end{subfigure}
     \hfill
    \caption{Visualization of the decoded image under different conditions. (a) is the original frame. (b) is the reconstructed frames using the pretrained autoencoder when there is no loss. (c)-(d) are the reconstructed frames using the pretrained model and \newae when 50\% loss happens in 3 consecutive frames, respectively.}
    \label{fig:example-image}
\end{figure*}

Figure~\ref{fig:example-image} gives a concrete visualization of how well \newae performs under packet losses. When there is no loss, both the pretrained autoencoder and \newae can reconstruct the frame with a decent quality (Figure~\ref{fig:example-pretrained-noloss}) with 0.9952 and 0.9943 SSIM, respectively. When 50\% loss are applied to the last 3 consecutive frames, the pretrained autoencoder model fails to reconstruct the original frame with a good quality: annoying distortions exists in the decoded frame (at the bottom of the glass container) and SSIM decreased to 0.9630. However, under the same loss, \newae can have a much higher SSIM (0.9830) and the distortions are much less visible (Figure~\ref{fig:example-ours-withloss}) comparing to the pretrained model.

%\section{Sanity check of traditional codecs}

%\section{Validation of implemented rate control logic}

%\section{Per-packet entropy encoding}

% \section{Test video characteristics}

% \jc{
% \begin{itemize}
% \item check bpp on x-axis
% \item check legend (ae, \name, etc)
% \item check function name ambiguity
% \end{itemize}
% }

%%%%%%%%%%%%%%%%%%%%%%%%%%%%%%%%%%%%%%%%%%%%%%%%%%%%%%%%%%%%%%%%%%%%%%%%%%%%%%%%
\end{document}